\iffalse Before sending further, pass this check list:
  [ ] find and resolve all "??" and "\ifnote"
  [ ] Spell-check
  [ ] order of using/defining abbreviations
  [ ] citation order (use RefTest)
  [ ] figure order & reference order
  [ ] Check LaTeX  output files (*.log) for warnings
  [ ] Check BibTeX output (screen) for warnings
  [ ] update PACS
  [ ] decide on color figures
  [ ] Re-read paper in the morning

After completing this list, if you made at least one correction,
re-do this check-list from the beginning, until no corrections will
be done.\fi

\documentclass[amsmath,amssymb,prb,preprintnumbers,showpacs,superscriptaddress, twocolumn]{revtex4}

\usepackage{graphicx}
\usepackage{dcolumn}    
\usepackage{bm}         
\usepackage{color}      
\usepackage{ulem}       

\newcommand{\lapprox}{{\scriptscriptstyle\stackrel{<}{\sim}}}

\newif\ifnote

\newif\iffig

\figtrue                     %

\begin{document}


\title{Transport, magnetic, and structural properties of
La$_{0.7}$Ce$_{0.3}$MnO$_3$ thin films. Evidence for hole-doping}

\author{R.~Werner}
\affiliation{%
  Physikalisches Institut -- Experimentalphysik II,
  Universit\"{a}t T\"{u}bingen,
  Auf der Morgenstelle 14,
  72076 T\"{u}bingen, Germany
}
\author{C.~Raisch}
\affiliation{%
  Physikalische Chemie,
  Universit\"{a}t T\"{u}bingen,
  Auf der Morgenstelle 14,
  72076 T\"{u}bingen, Germany
}
\author{V.~Leca}
\altaffiliation{Permanent address: University Politehnica Bucharest,
Faculty of Applied Chemistry and Materials Science, Science and
Engineering of Oxide Materials and Nanotechnology Department,
Gheorghe Polizu Street, no. 1-7, 011061, Bucharest, Romania}
\affiliation{%
  Physikalisches Institut -- Experimentalphysik II,
  Universit\"{a}t T\"{u}bingen,
  Auf der Morgenstelle 14,
  72076 T\"{u}bingen, Germany
}
\author{V.~Ion}
\altaffiliation{Permanent address: National Institute for Lasers,
Plasma and Radiation Physics, 409 Atomistilor Street, PO-Box MG-16,
077125 Magurele - Bucharest, Romania}
\author{S.~Bals}
\author{G.~Van Tendeloo}
\affiliation{%
  EMAT
  University of Antwerp,
  Groenenborgerlaan 171
  B-2020 Antwerp, Belgium
}
\author{T.~Chass\'{e}}
\affiliation{%
  Physikalische Chemie,
  Universit\"{a}t T\"{u}bingen,
  Auf der Morgenstelle 14,
  72076 T\"{u}bingen, Germany
}
\author{R.~Kleiner}
\author{D.~Koelle}
\email{koelle@uni-tuebingen.de}
\affiliation{%
  Physikalisches Institut -- Experimentalphysik II,
  Universit\"{a}t T\"{u}bingen,
  Auf der Morgenstelle 14,
  72076 T\"{u}bingen, Germany
}

\date{\today}

\begin{abstract}
Cerium-doped manganite thin films were grown epitaxially by pulsed
laser deposition at $720\,^\circ$C and oxygen pressure
$p_{O_2}=1-25\,$Pa and were subjected to different annealing steps.
According to x-ray diffraction (XRD) data, the formation of CeO$_2$
as a secondary phase could be avoided for $p_{O_2}\ge 8\,$Pa.
However, transmission electron microscopy shows the presence of
CeO$_2$ nanoclusters, even in those films which appear to be single
phase in XRD.
With O$_2$ annealing, the metal-to-insulator transition temperature
increases, while the saturation magnetization decreases and stays
well below the theoretical value for electron-doped
La$_{0.7}$Ce$_{0.3}$MnO$_3$ with mixed Mn$^{3+}$/Mn$^{2+}$ valences.
The same trend is observed with decreasing film thickness from 100 to
20\,nm, indicating a higher oxygen content for thinner films.
Hall measurements on a film which shows a metal-to-insulator
transition clearly reveal holes as dominating charge carriers.
Combining data from x-ray photoemission spectroscopy, for
determination of the oxygen content, and x-ray absorption
spectroscopy (XAS), for determination of the hole concentration and
cation valences, we find that with increasing oxygen content the hole
concentration increases and Mn valences are shifted from 2+ to 4+.
The dominating Mn valences in the films are Mn$^{3+}$ and Mn$^{4+}$,
and only a small amount of Mn$^{2+}$ ions can be observed by XAS.
Mn$^{2+}$ and Ce$^{4+}$ XAS signals obtained in surface-sensitive
total electron yield mode are strongly reduced in the bulk-sensitive
fluorescence mode, which indicates hole-doping in the bulk for those
films which do show a metal-to-insulator transition.
%
\end{abstract}

\pacs{75.47.Lx,72.60.+g,71.30.+h,61.05.cj,68.37.Lp,81.15.Fg}

\maketitle

\section{Introduction}
\label{Sec:Introduction}

Hole-doped manganese perovskite oxides La$_{1-x}A_x$MnO$_3$, where
$A$ is a divalent alkaline earth metal, have been intensively studied
over the last years due to the interesting interplay between charge,
spin, orbital and structural degrees of freedom.\cite{Imada98,Coey99,
Salamon01}
Without doping, LaMnO$_3$ is an antiferromagnetic insulator due to
the super-exchange between the Mn$^{3+}$ ions.\cite{Millis98}
In the hole-doped manganites, the divalent ion introduces holes by
changing some Mn valences from Mn$^{3+}$ to Mn$^{4+}$.
The properties of the hole-doped manganites are determined by the
interplay of Hund´s rule coupling and the Jahn-Teller distortion of
the Mn$^{3+}$ ions.\cite{Millis95}
Their behavior can be qualitatively described by the double-exchange
model,\cite{Zener51, Anderson55} describing the interaction between
manganese ions with mixed valences (Mn$^{3+}$ and Mn$^{4+}$).
The strong spin-charge coupling via the double-exchange interaction
explains the correlation between the metal-to-insulator (MI) and
ferromagnet-to-paramagnet (FP) transition.
Close to the MI transition temperature $T_{MI}$ an external magnetic
field can reduce the spin disorder and therefore enhance the electron
hopping between the manganese ions with mixed valences.
This results in a large resistivity drop, called colossal
magnetoresistance.\cite{Jonker50}

By substitution of La with a tetravalent ion, like Ce,\cite{Mandal97,
Gebhardt99, Ganguly00} Sn,\cite{Li99a} or Te,\cite{Tan03a} instead of
a divalent one, some of the Mn$^{3+}$ ions become Mn$^{2+}$ with
electronic structure t$^3_{2g}$e$^2_g$ (compared to the
t$^3_{2g}$e$^1_g$ electronic structure for Mn$^{3+}$).
Hence, an extra electron may be induced in the e$_g$-band.
Since Mn$^{2+}$ is a non-Jahn-Teller ion, like Mn$^{4+}$, one might
expect a similar magnetic interaction between the Mn$^{3+}$ and
Mn$^{2+}$ ions as for the well known hole-doped case.\cite{Mitra03a}

The first attempts to achieve electron-doping by substituting La with
Ce were reported by Mandal and Das.\cite{Mandal97}
However, they found hole-doping in their bulk samples.
Later on, it was revealed that the bulk samples are a multiphase
mixture which leads to the hole-doped behavior.\cite{Ganguly00,
Philip99}
Single phase {La$_{0.7}$Ce$_{0.3}$MnO$_3$} (LCeMO) thin films have
been prepared without any CeO$_2$ impurities \cite{Mitra01a,
Raychaudhuri99} regarding x-ray diffraction (XRD) data.
The films showed FP and MI transitions similar to the hole-doped
manganites.
Surface-sensitive X-ray photoemission spectroscopy revealed the
existence of Mn$^{2+}$ and Mn$^{3+}$ valences,\cite{Mitra03a, Han04}
which was interpreted as evidence of electron-doping.
However, Hall measurements and thermopower measurements on comparable
samples showed a hole-type character.
\cite{Wang06,Zhao00,Yanagida04,Yanagida05}
By Ganguly {\it et al.}~\cite{Ganguly00} it was further questioned
whether LaMnO$_3$ accepts Ce-doping at all.
Those authors questioned the reports on single phase LCeMO-films and
claimed the presence of multi-phase mixtures, consisting of hole
doped La-deficient phases with cerium oxide inclusions.
Certainly, the existence of electron-doped manganites could enable
new types of spintronic devices, such as $p-n$ junctions based on
doped manganites.\cite{Mitra01}
This motivates further research in order to improve understanding of
the basic properties of those materials.

In this paper we present the results of studies on transport,
magnetic and structural properties of LCeMO thin films grown by
pulsed laser deposition (PLD) and their dependence on deposition
parameters, annealing procedures and film thickness.
We combine a variety of different characterization techniques in
order to clarify the nature of the FP and MI transition in our LCeMO
thin films.

\section{Experimental Details}
\label{Sec:Experiment}

A commercially available stoichiometric polycrystalline
La$_{0.7}$Ce$_{0.3}$MnO$_3$ target was used for thin film growth by
PLD on (001) SrTiO$_3$ (STO) substrates (unless stated otherwise).
The target was ablated by using a KrF ($\lambda$ = 248 nm) excimer
laser at a repetition rate of $2-5\,$Hz.
The energy density on the target was $E_d=2\,{\rm J/cm}^2$, while the
substrate temperature during deposition was kept at
$T_s=720\,^{\circ}$C for all films for which data are presented
below, except for sample K  with slightly lower $T_s$ and $E_d$
(cf.~Tab.~\ref{tab:overview}).
The oxygen pressure $p_{O_2}$ during film growth was varied in the
1--25\,Pa range with the aim of yielding single phase films with
optimum morphology.
We used a relatively low deposition pressure as compared to some
literature data \cite{Mitra03a, Wang06,Mitra01,Chang04} in order to
avoid over-oxygenation of the films.
This is important, as it is known that perovskite rare-earth
manganites can accept a large excess of oxygen via the formation of
cation vacancies, inducing hole-doping in the parent compound
LaMnO$_3$.\cite{Toepfer97}
In-situ high-pressure reflection high energy electron diffraction
(RHEED) was used to monitor the growth mode and film thickness.
After deposition, most of the films were in-situ annealed for 1\,h at
$T=700\,^{\circ}$C and $p_{O_2}=1\,$bar and then cooled down with
10$^{\circ}$C per minute.
In the following, those samples will be called ''in-situ annealed``
films, in contrast to the ''as-deposited`` films which were just
cooled down to room temperature under deposition pressure.
Some of the samples have been additionally annealed ex-situ at
$p_{O_2}=1\,$\,bar in one or two steps ($1^{\rm st}$ step at
$700\,^{\circ}$C; $2^{\rm nd}$ step at $750\,^{\circ}$C; each step
for one hour).
Table \ref{tab:overview} summarizes the fabrication conditions and
some characteristics of the LCeMO films described below.

\begin{table}
\begin{tabular}{ccccccc}
\hline\hline
\rule{0mm}{4mm}
\#              & $p_{O_2}$ (Pa)    & \multicolumn{2}{c}{annealing} & $d$ (nm)  & $c$-axis ($\AA$)  & $T_{MI}$ (K)  \\
                &                   &  in-situ  & ex-situ           &           &                   &               \\
\hline
\vspace{1mm}A   &   1               &   no      &   no              &   100     &   3.921           &    --         \\
B1              &                   &           &   no              &           &   3.905           &   175         \\
\vspace{1mm}B2  & \raisebox{1.5ex}[-1.5ex]3 & \raisebox{1.5ex}[-1.5ex]{no} & $1\times$ & \raisebox{1.5ex}[-1.5ex]{90} & -- & 250\\
\vspace{1mm}C   &   8               & yes       &   no              &   100     &   3.897           &   190         \\
\vspace{1mm}D   &   25              &   no      &   no              &   100     &   3.880           &   180         \\
E1              &                   &           &   no              &           &   3.894           &   210         \\
E2              &   8               & yes       &   $1\times$       &   65      &   3.887           &   216         \\
\vspace{1mm}E3  &                   &           &   $2\times$       &           &   3.872           &   230         \\
\vspace{1mm}F   &   8               & yes       &   no              &   40      &   3.879           &   223         \\
\vspace{1mm}G   &   8               & yes       &   no              &   20      &   3.870           &   232         \\
\vspace{1mm}H   &   3               & $(^*)$    &   no              &   100     &   3.876           &   260         \\
K               & 3$(^{**})$        & no        &   no              &   50      &   3.894           &   180         \\
\hline\hline
\end{tabular}
\newline
$(^*)$ Cooled in 1\,bar O$_2$ without 1\,hour in-situ annealing
\newline
$(^{**})$ deposited at $T_s=700\,^{\circ}$C with $E_d=1.75\,{\rm J/cm}^2$
\caption{\ifnote{\sf\textcolor{red}{[tab:overview]\;}}\fi
Deposition pressure $p_{O_2}$, annealing conditions, thickness,
$c$-axis and transition temperature $T_{MI}$ of LCeMO films
investigated.}
\label{tab:overview}
\end{table}

The surface morphology was checked by atomic force microscopy (AFM)
in contact mode.
The crystal structure of the films was characterized by XRD and by
high-resolution (HR) transmission electron microscopy (TEM).
Transport properties were measured with a four probe technique, and a
superconducting quantum interference device (SQUID) magnetometer was
used to determine the magnetic properties of the samples.
Hall measurements were performed in order to obtain information on
the dominating type of charge carriers,
and x-ray photoemission spectroscopy (XPS) was performed in order to
obtain information on the oxygen content of different samples.
The valences of the manganese and cerium ions were evaluated by x-ray
absorption spectroscopy (XAS).
XAS measurements in surface-sensitive total electron yield (TEY) mode
and bulk sensitive fluorescence yield (FY) mode were carried out at
the WERA dipole beamline (ANKA, Karlsruhe, Germany)  with typical
energy resolutions set between 100 and 400\,meV.

\section{Structural Analysis}

\begin{figure}
\iffig
\centering
\includegraphics[width=0.45\textwidth]{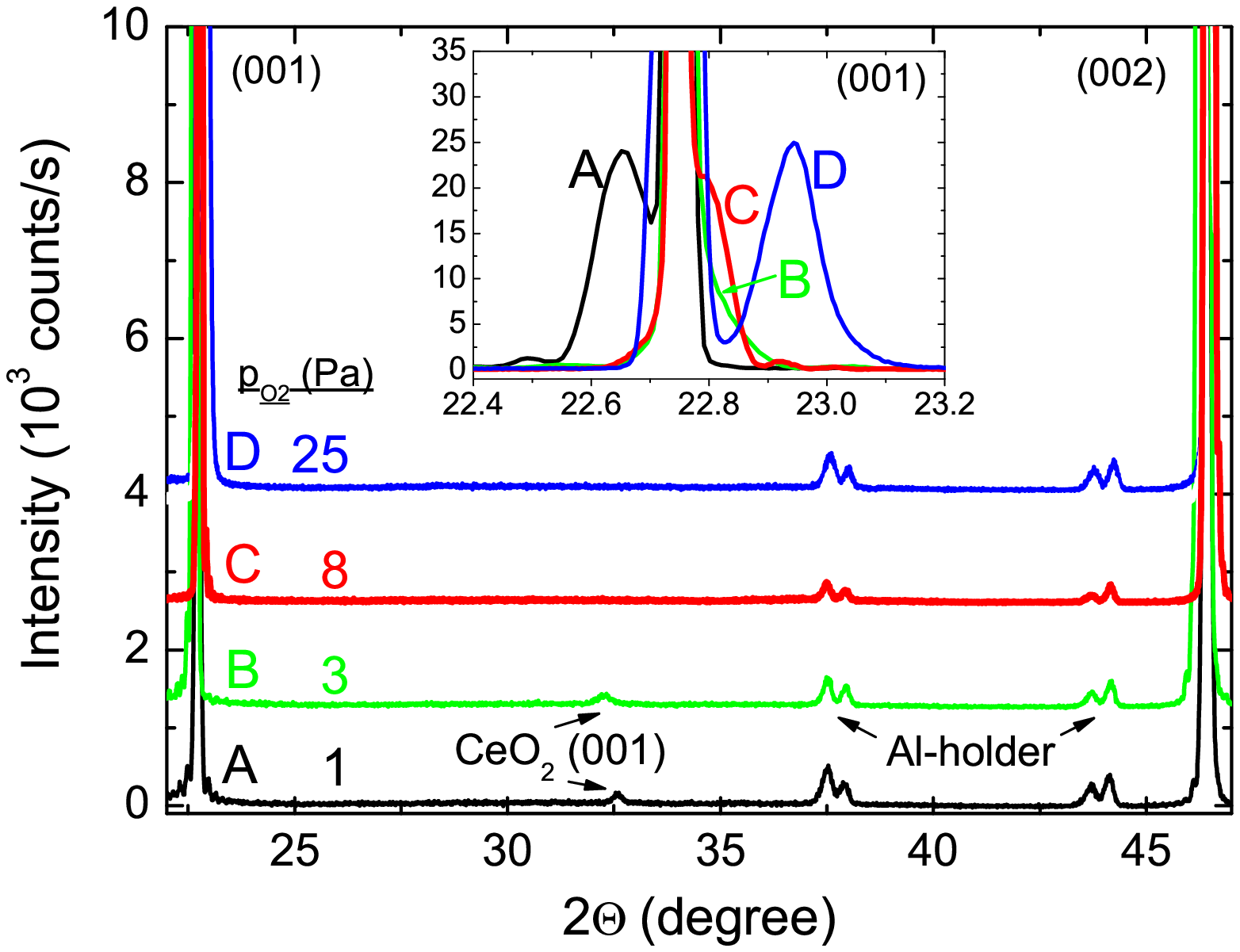}
\fi
\ifnote{\sf\textcolor{red}{[filename: XRDCeO2]\;}}\fi
\caption{\ifnote{\sf\textcolor{red}{[label: XRDCeO2]\;}}\fi
(Color online)
XRD patterns of samples grown under different deposition pressures:
$p_{O_2}=1,\,3,\,8$ and $25\,$Pa for sample A, B, C and D, respectively.
CeO$_2$ can be identified in samples A and B.
XRD scans are offset for clarity.
The inset shows a detailed view around the (001) substrate peak
including the (001) film peaks.}
\label{XRDCeO2}
\end{figure}

Figure \ref{XRDCeO2} shows the XRD $\Theta-2\Theta$ scans of four
LCeMO thin films A, B, C, D (with similar thickness $d=$ 90--100\,nm)
grown under different oxygen pressure $p_{O_2}=$ 1, 3, 8 and 25\,Pa,
respectively.
Sample C was in-situ annealed while the other samples were
''as-deposited`` films.
According to the XRD data shown in Fig.~\ref{XRDCeO2}, single phase
LCeMO films were obtained for $p_{O_2}\ge 8\,$ Pa (samples C and D).
For a lower deposition pressure, impurity peaks of CeO$_2$ appear
(sample A and B).
The substrate temperature $T_s$ also played a crucial role for the
phase stability of the LCeMO films.
By increasing $T_s$ up to $800\,^{\circ}$C, CeO$_2$ also appears for
deposition pressures $p_{O_2}\ge 8\,$Pa.
Such a behavior was also observed by Chang {\it et
al.}.\cite{Chang04}
As shown in the inset of Fig.~\ref{XRDCeO2}, the $c$-axis decreases
with increasing deposition pressure $p_{O_2}$.
This can be explained by a decreasing concentration of oxygen
vacancies with increasing $p_{O_2}$, as it is well known that oxygen
vacancies tend to expand the lattice constants.\cite{Murugavel03}

\begin{figure}
\iffig
\centering
\includegraphics[width=0.43\textwidth]{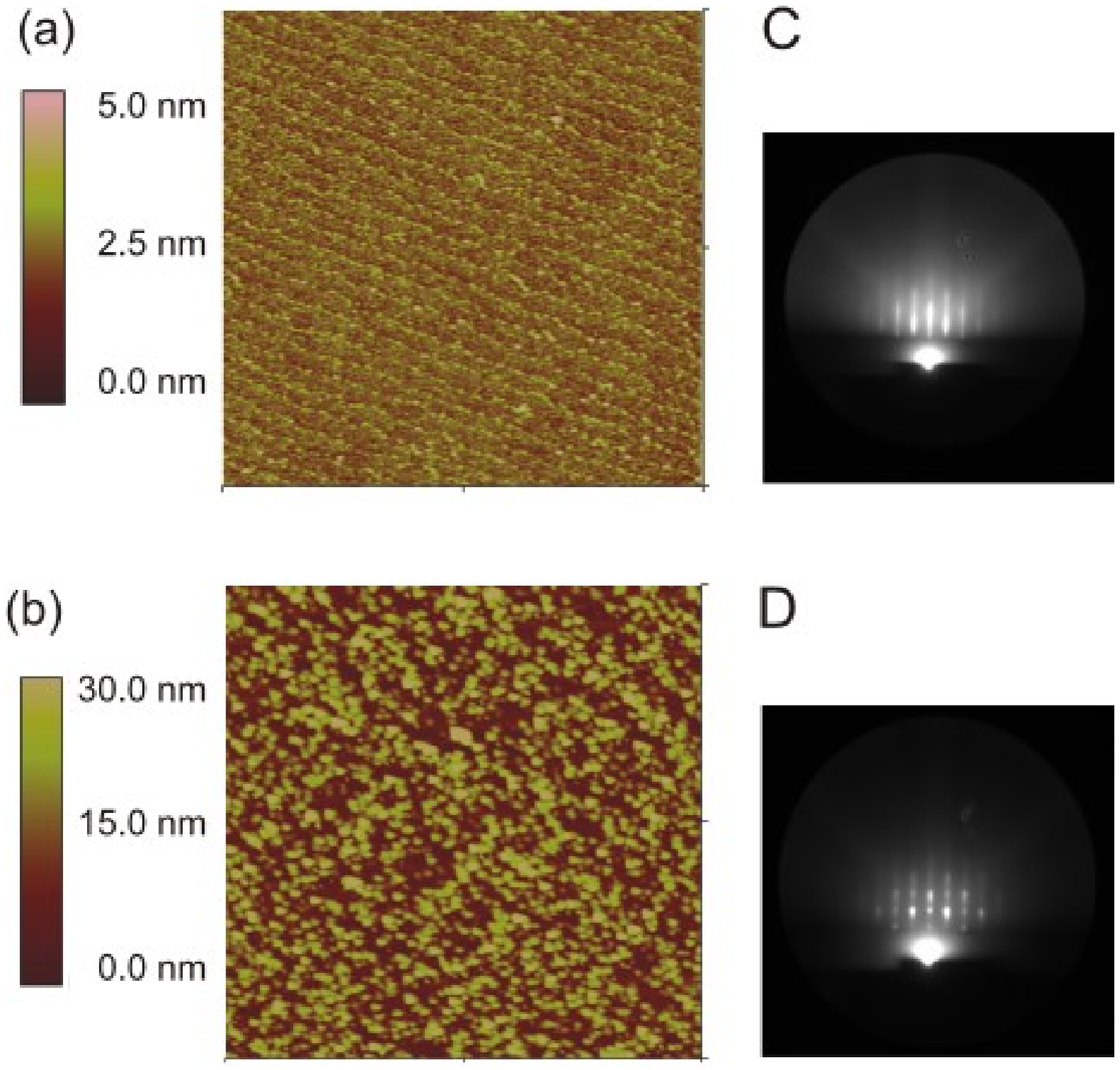}
\fi
\ifnote{\sf\textcolor{red}{[AFM-RHEED]\;}}\fi
\caption{\ifnote{\sf\textcolor{red}{[AFM-RHEED]\;}}\fi
(Color online)
AFM images (left; frame size $5\times 5\,\mu{\rm m}^2$)
and RHEED images (right) of 100\,nm thick films:
(a) sample C, grown at $p_{O_2}=8\,$Pa, and
(b) sample D, grown at $p_{O_2}=25\,$Pa.}
\label{AFM-RHEED}
\end{figure}

The surface roughness of the films depends strongly on deposition
pressure, as shown by AFM and RHEED images on 100\,nm thick films in
Fig.~\ref{AFM-RHEED} for (a) sample C ($p_{O_2}=8\,$Pa) with an rms
roughness of 0.35\,nm and (b) sample D ($p_{O_2}=25\,$Pa;), with a
much larger rms value of 2.15\,nm.
The RHEED images show strong streaky patterns for the film deposited
at $p_{O_2}=8\,$Pa [Fig.~\ref{AFM-RHEED}(a) right], an indication of
an atomically flat surface, while for higher deposition pressure
[here $p_{O_2}=25\,$Pa; Fig.~\ref{AFM-RHEED}(b) right] an increased
surface roughness results in a combination of weaker streaks,
together with the formation of a 3D RHEED pattern as a result of
island growth.
We note that sample C has an extremely smooth surface, showing
unit-cell high terrace steps in the AFM image
[c.f.~Fig.~\ref{AFM-RHEED}(a) left], which is quite unusual for such
a thick LCeMO film.
A similar morphology as for sample C was observed for all films
deposited at an oxygen pressure in the range of 1-8\,Pa.
For those conditions the films followed a 2D growth mode, as
suggested by the RHEED and AFM data.
Increasing the deposition pressure resulted in an increased step
density during growth due to lower surface mobility, with the
formation of 3D islands.
Altogether, we found that $p_{O_2}=8\,$Pa was the optimum pressure
for growing films without measurable CeO$_2$ concentration, as
detected by XRD, and good surface morphology (rms roughness below
0.4\,nm).

In order to evaluate the relation between CeO$_2$ formation and the
substrate induced strain,
50\,nm thick LCeMO films were deposited on (001) STO, (110) NdGaO$_3$
and (001) NdGaO$_3$ substrates in the same deposition
run.\footnote{From the bulk values for the LCeMO lattice constants
$a=3.821\,\AA$ and $b=3.902\,\AA$ \cite{Chang04} one obtains an
in-plane lattice mismatch ranging from -2\,\% (tensile strain) to
+1\,\% (compressive strain) for the different substrates used here.}
Here, we used a deposition pressure $p_{O_2}=3\,$Pa, in order to
obtain a measurable amount of CeO$_2$.
The XRD data showed no discernible difference in the amount of
CeO$_2$ for the different substrates.

The growth and phase stability of some complex oxide materials may
depend on the type of termination layer of the
substrate.\cite{Huijbregtse01}
Therefore, we have grown several LCeMO films on (001) STO substrates
with different termination (either SrO or TiO$_2$) in order to
determine whether the substrate termination influences the
microstructure of the films.
The SrO terminated substrates were obtained by annealing at
$950\,^{\circ}$C, for 1\,h in an oxygen flow, while the
TiO$_2$-terminated STO substrates were obtained by chemical etching
in a BHF solution, following the procedure described in
Ref.~[\onlinecite{Koster98}].
The results showed no correlation between the substrate termination
and the CeO$_2$ impurity phase formation.
These results suggest that, for the conditions used in this study,
the level of strain and the type of substrate termination do not have
an important effect on the phase stability in the LCeMO system and
that, most probably, the deposition conditions (in particular $T_s$
and $p_{O_2}$) are the determining factors.

\begin{figure}
\iffig
\includegraphics[width=0.35\textwidth]{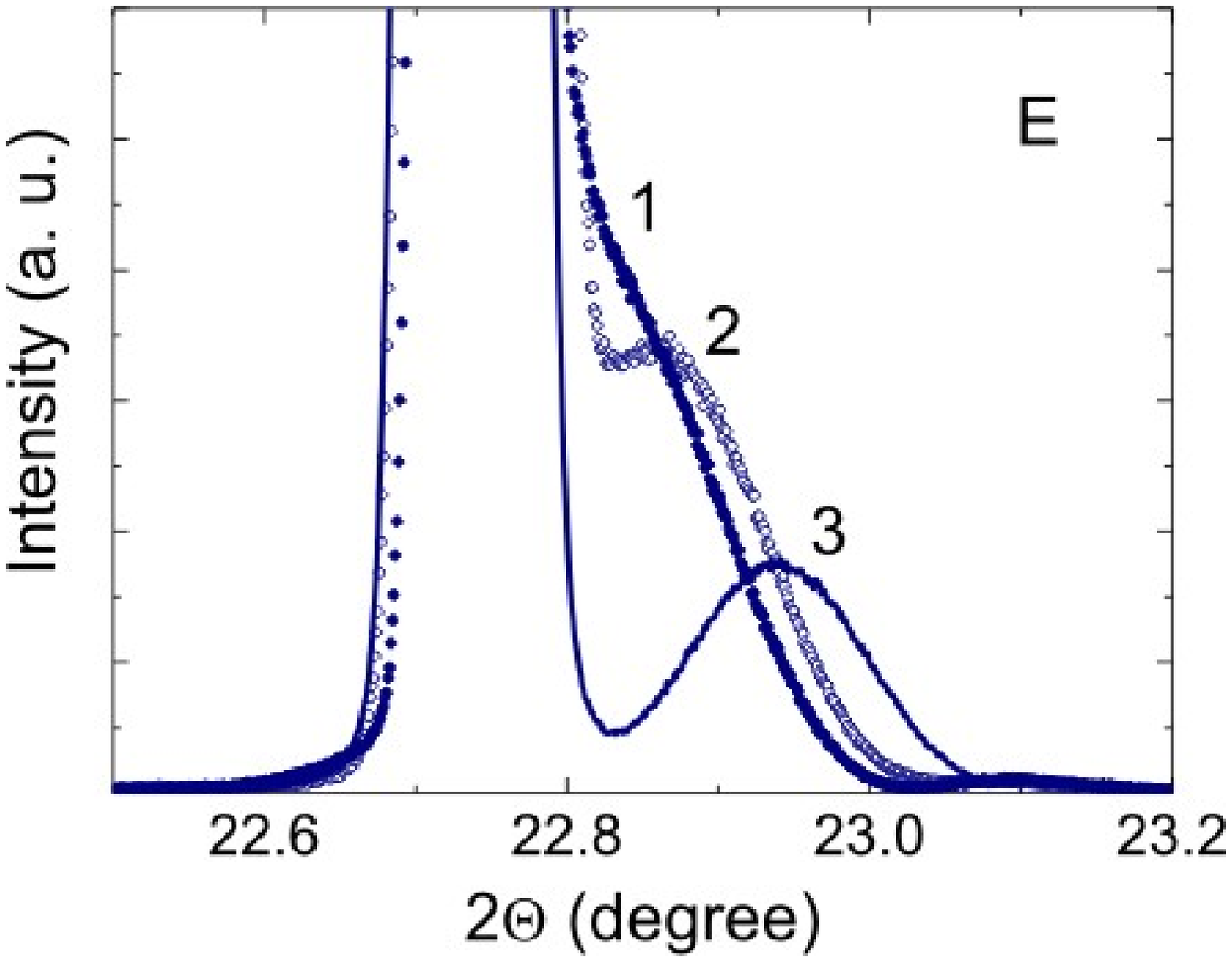}
\fi
\ifnote{\sf\textcolor{red}{[XRDrr33annealing]\;}}\fi
\caption{\ifnote{\sf\textcolor{red}{[annealing]\;}}\fi
(Color online)
XRD pattern at the (001) peak for sample E, showing the evolution of
the $c$-axis with ex-situ annealing steps: after in-situ annealing
(1), first (2) and second (3) ex-situ annealing.}
\label{annealing}
\end{figure}

Figure \ref{annealing} shows the evolution of the $c$-axis with
additional ex-situ annealing steps as obtained from XRD data for the
(001) peak on sample E.
As a result, the $c$-axis decreased from $c=3.894\,${\AA} to
$c=3.872\,${\AA}.
As the $a$- and $b$-axis bulk values for LCeMO are smaller than the
ones of the STO substrate, the observed shrinking of the $c$-axis
cannot be related to strain relaxation effects (which would increase
$c$), but most probably to the incorporation of extra oxygen in the
film.
As another result of the annealing experiments, we did not find a
correlation between ex-situ annealing and the CeO$_2$ concentration
in our films.
This is in contrast to the observations presented by Yanagida {\it et
al.}\cite{Yanagida04} and Chang {\it et al.};\cite{Chang04} however,
in their work, much longer annealing times (up to 10 hours) have been
used.
In our case, samples without secondary phase stayed single phase
regarding the XRD data.
However, while XRD data indicate that films deposited at 8-25\,Pa
O$_2$ are single phase, HRTEM analysis showed evidence for phase
separation even in these samples.
The results of the microstructural TEM analysis are discussed in the
following.

\section{TEM}

To obtain a better understanding on the relation between the
microstructure and the physical properties of our LCeMO thin films, a
few samples grown at different oxygen pressure were selected for TEM
analysis.
Here, we show results obtained from two films: sample E
prepared at $p_{O_2}=8\,$Pa, which appears single phase at
XRD, and sample K prepared at $p_{O_2}=3\,$Pa,
containing CeO$_2$ as secondary phase.
TEM studies were carried out using a JEOL 4000EX microscope operated
at 400\,kV.
The instrument has a point-to-point resolution of 0.17\,nm.
Planview TEM specimens were prepared by mechanical polishing of the
samples down to a thickness of $30\,\mu$m, followed by Ar ion-milling
at grazing incidence to reach electron transparency.

\begin{figure}
\iffig
\centering
\includegraphics[width=0.45\textwidth]{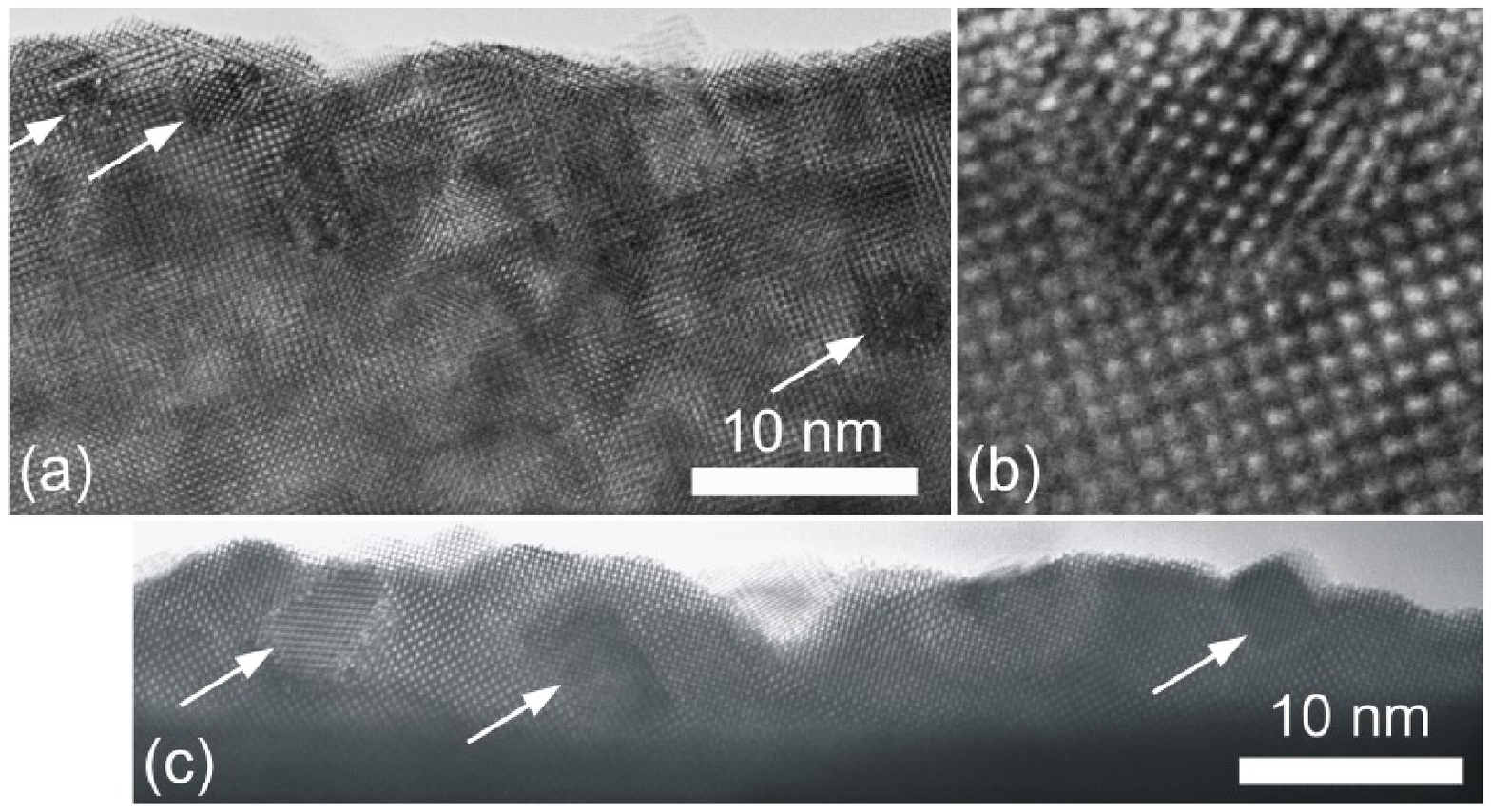}
\fi
\ifnote{\sf\textcolor{red}{[TEM]\;}}\fi
\caption{\ifnote{\sf\textcolor{red}{[TEM]\;}}\fi
(a) Planview HRTEM images of (a) sample E
grown at 8\,Pa O$_2$;
arrows indicate the CeO$_2$ inclusions.
An example of an inclusion [cf.~left arrow in (a)]
is shown in more detail in (b).
(c) sample K grown at 8\,Pa O$_2$.}
\label{TEM}
\end{figure}

Figure \ref{TEM}(a) shows a HRTEM plan view image of the LCeMO
thin film grown at 8\,Pa O$_2$ (sample E).
Several CeO$_2$ nanoclusters are indicated by arrows.
A more detailed HRTEM image of one of the clusters is shown in
Fig.~\ref{TEM}(b).
Figure \ref{TEM}(c) shows a TEM plan view image of the LCeMO thin
film grown at 3\,Pa O$_2$ (sample K).
In this sample, a higher density of CeO$_2$ nanoclusters in
comparison to sample E is observed.
Furthermore, the size of the clusters is also larger (although still
within the nanometer region).
The interface between the CeO$_2$ nanoclusters and the matrix is
better defined in comparison to sample E.

HRTEM data for the analyzed samples prove the presence of CeO$_2$
nanoclusters in the perovskite matrix (LCeMO) and show that CeO$_2$
segregation in the 3\,Pa sample is larger than in the 8\,Pa sample.
In case of the 8\,Pa sample (and for another 25\,Pa film not shown)
the small total volume of CeO$_2$ clusters made them untraceable by
XRD.
As an important consequence, our TEM data show that even LCeMO films
which appear to be single phase from XRD data contain CeO$_2$
nanoclusters.
This observation is important, as it has been shown \cite{Yanagida04}
that the valence state of Mn in LCeMO is sensitive to the degree of
Ce segregation, which drives the valences from Mn$^{3+}$ to
Mn$^{4+}$, even in the presence of Ce$^{4+}$.

\section{Transport and magnetic properties}

\begin{figure}[b]
\iffig
\centering
\includegraphics[width=0.45\textwidth]{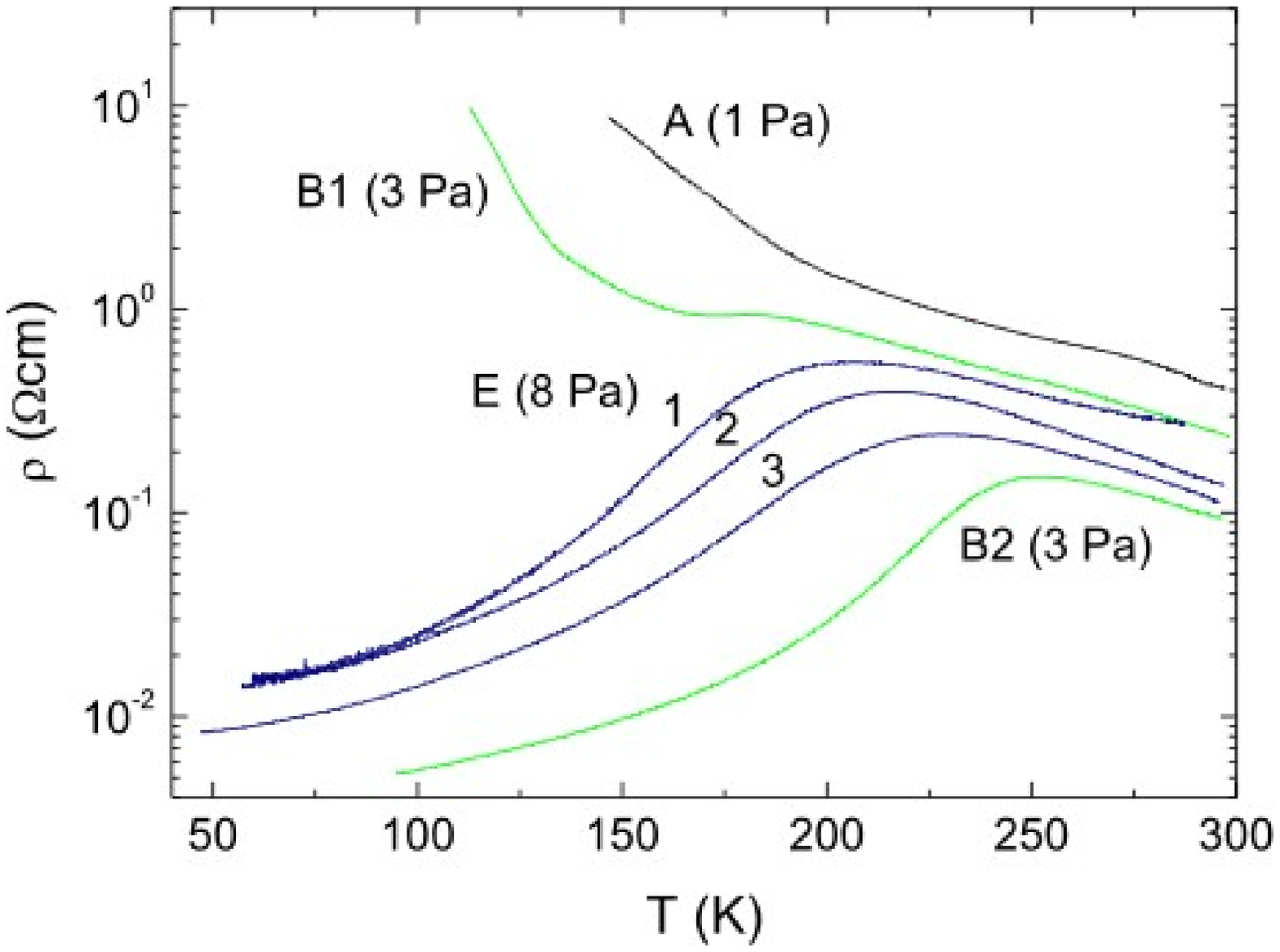}
\fi
\ifnote{\sf\textcolor{red}{[RT]\;}}\fi
\caption{\ifnote{\sf\textcolor{red}{[RT]\;}}\fi
(Color online)
Resistivity vs.~temperature for samples A, B and E (with deposition
pressure $p_{O_2}$ in parenthesis).
The behavior after ex-situ annealing is shown for sample B and E
(B1, E1: without ex-situ annealing;
B2, E2: after $1^{\rm st}$ ex-situ annealing step;
E3: after $2^{\rm nd}$ ex-situ annealing step).}
\label{RT}
\end{figure}

Figure \ref{RT} shows resistivity $\rho$ versus temperature $T$ for
samples A, B and E.
Sample A was ''as-deposited`` at $p_{O_2}=1\,$Pa and shows no
metal-to-insulator transition at all.
Due to its high resistivity we could not trace out $\rho(T)$ below
$T\approx 150\,$K.
Sample B, grown at 3\,Pa (also ''as-deposited``) shows a slight
indication of a metal-to-insulator transition, i.~e.~a maximum in
$\rho(T)$ at $T_{MI}= 175\,$K, with a strong increase in resistivity
at $T\lapprox 130\,$K, which can be explained by charge localization.
Sample E, grown at 8\,Pa (annealed in-situ) shows a transition at
$T_{MI}=210\,$K.

For sample E, the evolution of the $\rho(T)$ curves after two
annealing steps (c.~f.~Sec.\ref{Sec:Experiment}) is additionally
shown.
The $T_{MI}$ transition temperature increases to 230\,K, which is
accompanied by a decreasing resistivity, presumably due to an
increasing charge carrier density.
This observation is consistent with results obtained by Yanagida {\it
et al.}\cite{Yanagida04} and contradicts the picture of an
electron-doped manganite: Oxygen annealing should decrease the
concentration of Mn$^{2+}$ ions, hence, reduce the density of
electrons as charge carriers and therefore lower $T_{MI}$ and
increase resistivity.\cite{Wang06}
The annealing steps seem to create more Mn$^{4+}$ in the samples, and
the double-exchange between Mn$^{3+}$ and Mn$^{4+}$ gets stronger,
which leads to an increase of $T_{MI}$.
This interpretation is also supported by the results from
measurements of the saturation magnetization ($M_s$) and the
spectroscopic analysis, which will be discussed further below.

Figure \ref{RT} also shows that $T_{MI}$ of sample B increases more
drastically than sample E, even after only a single ex-situ annealing
step.
This might be due to the higher concentration of a secondary phase
(CeO$_2$) in sample B (c.~f.~Fig.~\ref{XRDCeO2}), which may favor
oxygen diffusion into the film due to crystal defects.

\begin{figure}[b]
\iffig
\centering
\includegraphics[width=0.35\textwidth]{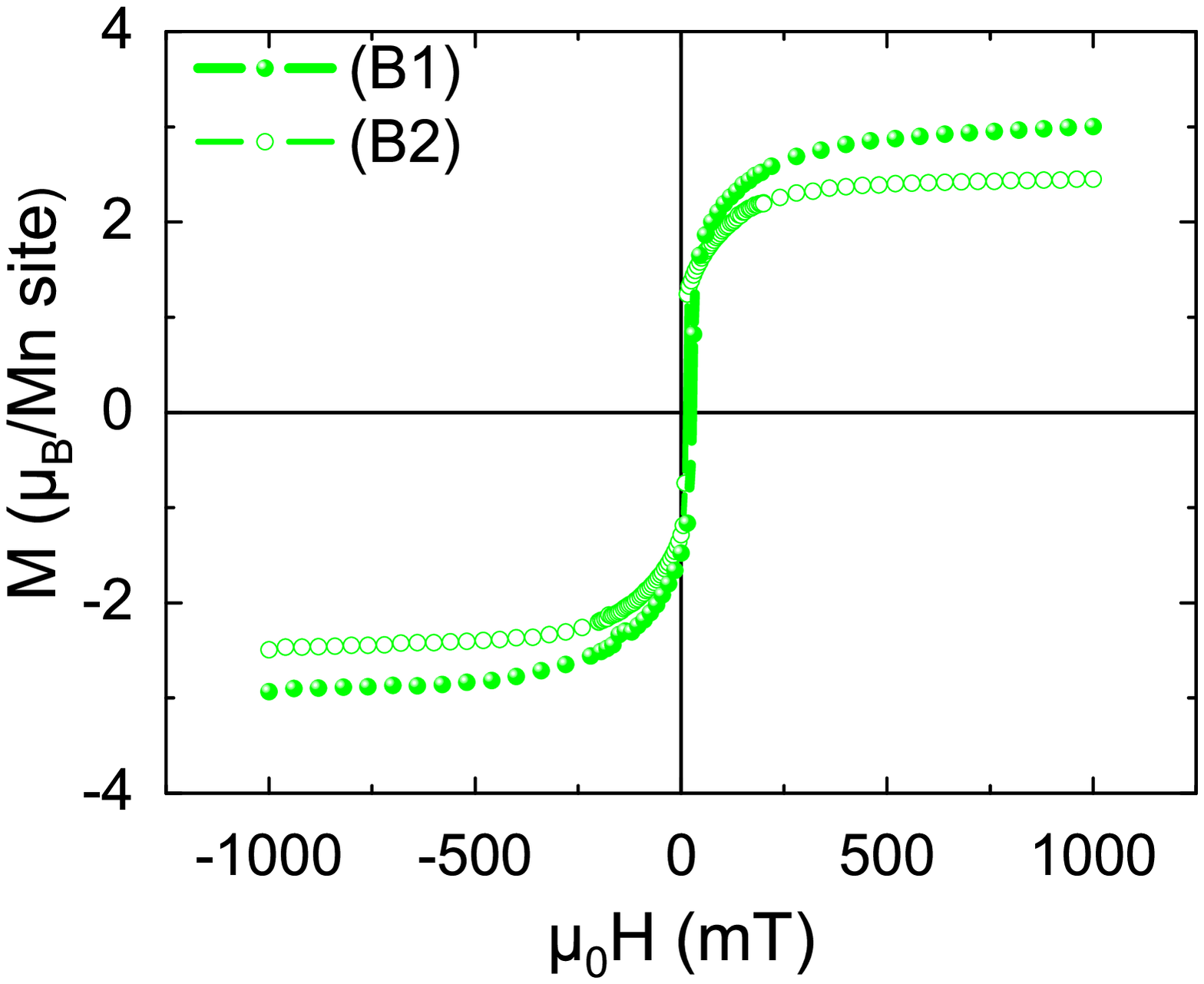}
\fi
\ifnote{\sf\textcolor{red}{[MvsH1]\;}}\fi
\caption{\ifnote{\sf\textcolor{red}{[MvsH1]\;}}\fi
(Color online)
Magnetization vs.~applied magnetic field at $T=20\,$K
for the as-grown ($p_{O_2}=3\,$Pa) sample B (B1)
and after ex-situ annealing (B2).}
\label{MvsH1}
\end{figure}

In Fig.~\ref{MvsH1} the magnetization $M$ (in units of $\mu_B/$Mn
site) vs.~applied field $\mu_0 H$ at $T=20\,$K is shown for sample B,
measured ''as grown`` (B1) and after ex-situ annealing (B2).
The ex-situ annealing step caused a decrease in the saturation
magnetization $M_s$, from 2.93 to $2.40\,\mu_B$/Mn-site, while
$T_{MI}$ increased from 175 to 250\,K.
With the magnetic moments $m=$5, 4 and $3\,\mu_B$ for Mn$^{2+}$,
Mn$^{3+}$ and Mn$^{4+}$, respectively, the theoretical value for the
saturation magnetization of electron-doped LCeMO is
$M_s=4.3\,\mu_B$/Mn-site.\cite{Zhang03}
Until now, this value has never been achieved.
However, for the hole-doped manganites, it is known that excess
oxygen increases the valences from Mn$^{3+}$ to Mn$^{4+}$, and
therefore decreases the magnetization.
Hence, the observed decrease in $M_s$ with oxygen annealing can be
explained by the decrease in Mn$^{2+}$ and concomitant increase in
Mn$^{4+}$ concentration.

\begin{figure}[b]
\iffig
\centering
\includegraphics[width=0.45\textwidth]{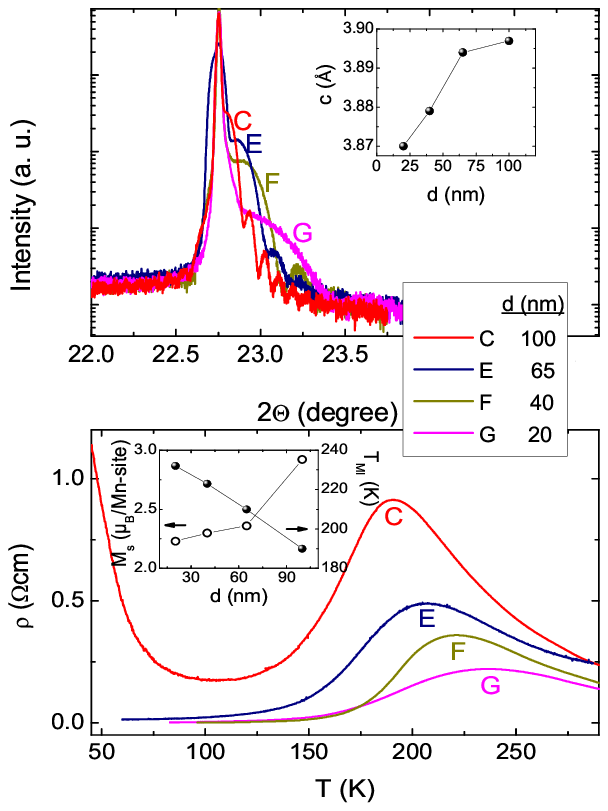}
\fi
\ifnote{\sf\textcolor{red}{[thickness]\;}}\fi
\caption{\ifnote{\sf\textcolor{red}{[thickness]\;}}\fi
(Color online)
Comparison of samples with different thickness $d$, grown under the
same deposition conditions ($p_{O_2}=8\,$Pa; in-situ annealed).
(a) XRD $\Theta - 2\Theta$ scans; the inset shows that the $c$-axis value increases with increasing $d$.
(b) Resistivity vs.~temperature; the inset shows that the transition temperature $T_{MI}$
decreases and the saturation magnetization $M_s$ (from $M(H)$ data; not shown) increases with increasing $d$.}
\label{thickness}
\end{figure}

In order to study the dependence of structural, transport and
magnetic properties on film thickness $d$, four samples (C, E, F, G
with $d$=100, 65, 40 and 20\,nm, respectively) were grown under the
same conditions, i.e., at $p_{O_2}=8\,$Pa with in-situ annealing.
The $\Theta-2\Theta$ XRD scans of the (001) peak in
Fig.~\ref{thickness}(a) show that with decreasing film thickness the
$c$-axis shrinks [see inset].
Assuming a fixed unit cell volume, this observation might be
explained by increasing tensile strain with decreasing $d$, as the
bulk in-plane lattice parameters of LCeMO are smaller than those for
the STO substrate.
However, as oxygen vacancies tend to expand the lattice parameters,
an increasing lack of oxygen with increasing $d$ has the same effect.
The transport properties shown in Fig.~\ref{thickness}(b) indicate
exactly this lack of oxygen with increasing film thickness.
Sample C, with largest $d$, shows again charge localization at low
$T$, while the thinnest film has the highest $T_{MI}$ [c.~f.~inset]
and lowest $\rho$.
>From magnetization measurements on samples C, E, F and G we also find
that $M_s$ increases with $d$ [c.~f.~inset in
Fig.~\ref{thickness}(b)].
The lowest saturation magnetization for the thinnest sample G is
another indication for the higher oxygen concentration compared to
the others.

\section{Hall measurements}

\begin{figure}[h]
\iffig
\centering
\includegraphics[width=0.35\textwidth]{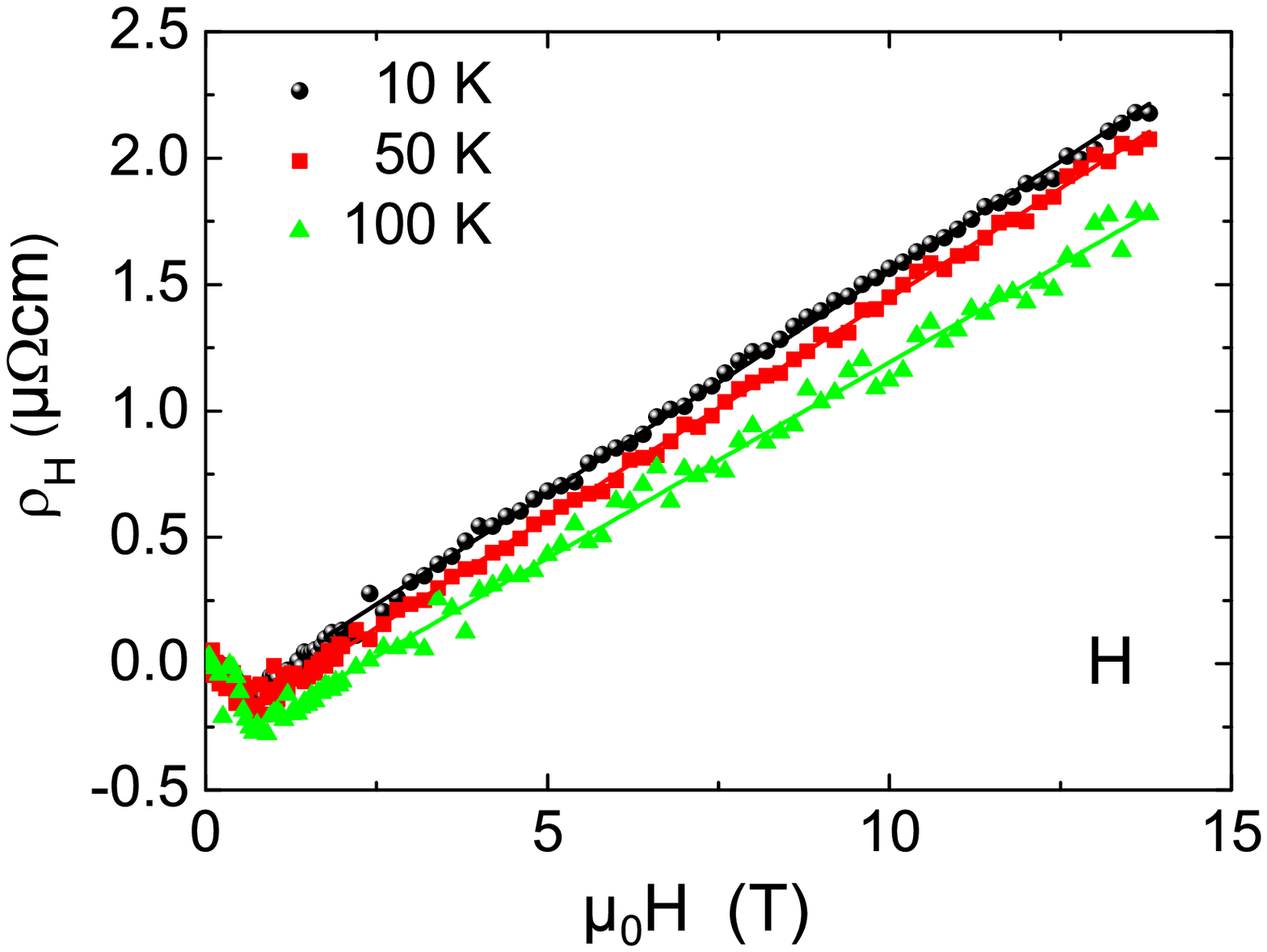}
\fi
\ifnote{\sf\textcolor{red}{[Hall]\;}}\fi
\caption{\ifnote{\sf\textcolor{red}{[Hall]\;}}\fi
(Color online)
Field dependence of the Hall resistivity of sample H.
The positive slope at high magnetic field identifies the majority of
the carriers to be holes.
The solid lines are linear fits to the high-field data.}
\label{Hall}
\end{figure}

In order to determine the type of majority charge carriers via the
Hall effect, we chose one of our films (sample H, $d=100\,$nm) which
was deposited at relatively low oxygen pressure ($p_{O_2}=3\,$Pa) and
cooled in 1\,bar, without an annealing step.
>From measurements of the longitudinal resistivity $\rho(T)$ of the
patterned film we find a clear MI transition with rather high
$T_{MI}=260\,$K.
The Hall resistivity $\rho_H$ was measured at $T=10$, 50 and 100\,K
in magnetic fields up to 14\,T.
The sign of the Hall voltage was carefully checked by using an
$n$-doped silicon reference sample.
Figure \ref{Hall} shows $\rho_H$ vs.~applied magnetic field $H$.
The drop of $\rho_H$ in the low-field range reflects the so-called
anomalous Hall Effect, $\rho_{aH}=R_{aH}\mu_0M$, which is due to spin
orbit interaction.\cite{Karplus54}
Here, $R_{aH}$ is the Hall coefficient for the anomalous Hall effect.
With further increasing field, the data show the expected linear
behavior of the normal Hall effect $\rho_{nH}=R_{nH}\mu_0H$ with Hall
coefficient $R_{nH}=1/ne$ and charge carrier density $n$.
The main feature in Fig.~\ref{Hall} is the positive slope
$\partial\rho_H/\partial H$ at high fields, which reveals the
majority of the carriers to be holes with $n=1.57$, 1.60 and
$1.78\times 10^{22}\,{\rm cm}^{-3}$, for $T=10$, 50 and 100\,K,
respectively.
This corresponds to 0.94--1.07 holes/Mn-site.
The observation of hole-doping is consistent with the results from
transport and magnetization measurements discussed above and also
with the spectroscopic analysis, which will be presented in the
following section.

\section{Spectroscopic Analysis }

X-ray Absorption Spectroscopy (XAS) was performed on LCeMO thin films
prepared under different conditions, in order to investigate the
relation between the manganese valences, the oxygen content and
transport and magnetic properties.
In total electron yield (TEY) detection mode only the uppermost 5 -
10\,nm are probed, depending on the electron escape depth, while in
fluorescence yield (FY) mode x-ray photons are detected.
They have typical attenuation lengths from 100\,nm (Ce M edge) to
200\,nm (O K and Mn L edge), thus giving insight into the bulk
structure of the samples.

Here we compare two films, D (as-deposited) and G (in-situ annealed),
which were deposited at different oxygen pressure $p_{O_2}=25\,$Pa
and 8\,Pa, respectively.
>From XPS measurements we find that the oxygen content of G is higher
than the one of D.
This shows that the higher deposition pressure (for sample D) is not
the key to higher oxygen concentration, but that annealing is most
relevant.
Sample G shows a MI transition at 232\,K
[c.~f.~Fig.~\ref{thickness}(b)],
while sample D shows a weak transition at 180\,K and charge
localization at lower temperatures.

\begin{figure}[b]
\iffig
\centering
\includegraphics[width=0.42\textwidth]{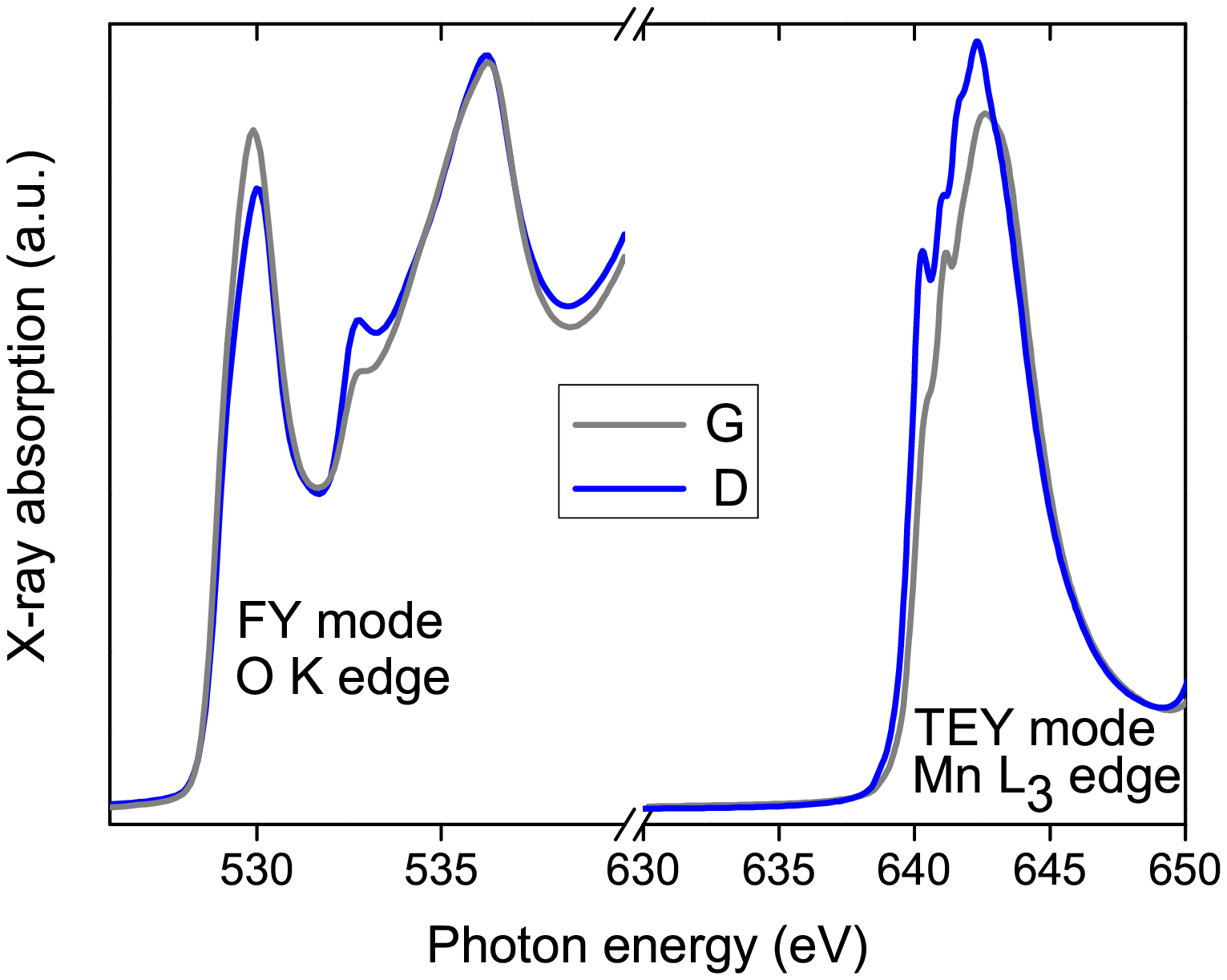}
\fi
\ifnote{\sf\textcolor{red}{[XAS-DvsG]\;}}\fi
\caption{\ifnote{\sf\textcolor{red}{[XAS-DvsG]\;}}\fi
(Color online)
XA spectra of samples D and G.
On the left side the oxygen K edge (FY mode) is shown
with the prepeak increasing with higher oxygen content.
The right side shows the manganese L$_3$ edge (TEY mode)
with different amounts of Mn$^{2+}$ for differently oxidized samples.}
\label{XAS-DvsG}
\end{figure}

A typical spectrum of the O K edge of LCeMO, measured in bulk
sensitive fluorescence yield (FY) mode, is seen in
Fig.~\ref{XAS-DvsG} (left).
The first structure at about 530\,eV arises from transitions from the
O1s level to states, which are commonly understood to be of mixed
Mn3d-O2p character and as being a measure of the hole
concentration.\cite{Abbate92,Manella05,Chang05}
In fact we found that this prepeak is stronger in sample G, i.e. the
more oxidized sample.
The second, rather broad and asymmetric feature at 532 to 537\,eV is
attributed to La5d (Ce), La4f (Ce) states hybridized with O2p states.
A third set of states (not shown here) is found at about 543\,eV and
is widely believed to derive from hybridization of O2p with higher
energy metal-states like Mn 4sp and La 6sp.\cite{Abbate92}

The corresponding spectrum at the Mn L edge taken in TEY detection
mode is shown in Fig.~\ref{XAS-DvsG} (right).
Both, the L$_3$ edge at 642\,eV and the L$_2$ edge at 653\,eV (not
shown here) are strongly broadened, indicating the presence of a
variety of valence states.
The most important feature is the shoulder at 640\,eV, which is a
clear indication of divalent Mn, as can be shown by a comparison with
XAS data from MnO.\cite{Nagel07}
In Fig.~\ref{XAS-DvsG} (right) this shoulder is more pronounced in
sample D, i.e., the less oxidized sample.
The relative spectral weight of this feature in combination with the
relative intensity of Mn3d-O2p states taken from the O K edge is
essential to explain the properties of the different samples.
A higher degree of oxidation leads to a higher relative spectral
weight of the O K prepeak and a lower amount of Mn$^{2+}$.
By introducing more oxygen, more holes are created and the manganese
valence is increased.
This finding is further supported by measurements on three additional
samples (not shown here), also showing the effect of film thickness,
oxygen pressure during growth and duration of post-growth annealing
in oxygen.

\begin{figure}
\iffig
\centering
\includegraphics[width=0.42\textwidth]{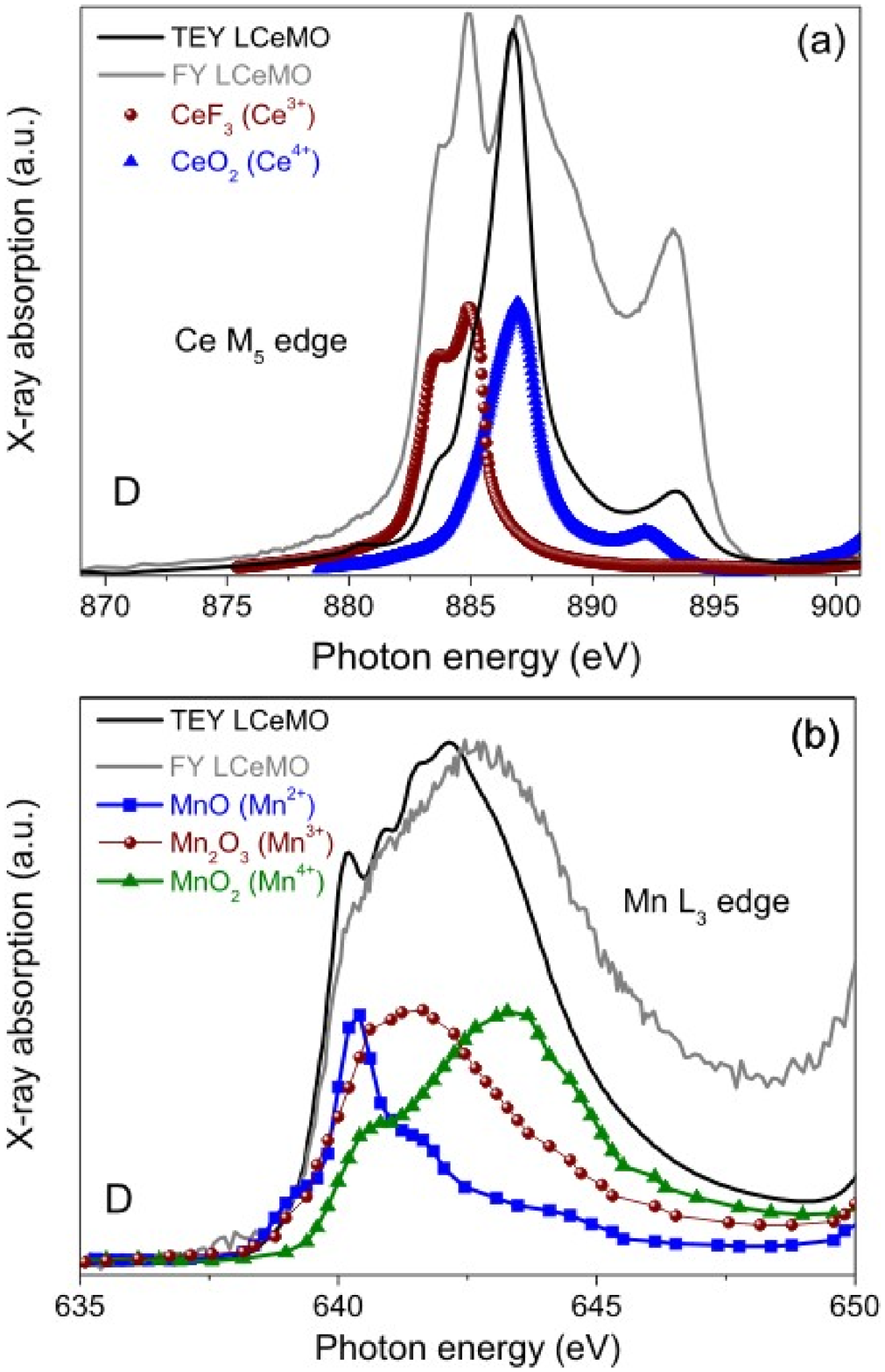}
\fi
\ifnote{\sf\textcolor{red}{[XAS-Ce-Mn]\;}}\fi
\caption{\ifnote{\sf\textcolor{red}{[XAS-Ce-Mn]\;}}\fi
(Color online)
XA spectra of sample D in TEY and FY mode (scaled to the TEY intensity)
at different absorption edges:
(a)
Cerium M$_5$ edge; reference spectra of CeO$_2$ and CeF$_3$
were added for comparison.
Please note the mixture of Ce$^{3+}$ and Ce$^{4+}$ in FY mode.
(b)
Manganese L$_3$ edge.
The FY data are self-absorption corrected following a procedure
by Ref.~[\onlinecite{Troeger92}].
Reference spectra of MnO (blue), Mn$_2$O$_3$ and MnO$_2$
were added for comparison.
Please note the missing Mn$^{2+}$ shoulder in FY mode.}
\label{XAS-Ce-Mn}
\end{figure}

The remaining issue is the oxidation state of the Ce ions, which is
important for the type of doping.
Looking at the Ce absorption M$_5$ edge both in surface-sensitive TEY
detection mode and bulk-sensitive FY mode, we found striking
differences in the spectral shapes of the measured spectra, as shown
in Fig.~\ref{XAS-Ce-Mn}(a).
Cerium reference data for CeO$_2$ and CeF$_3$ were taken from
Ref.~[\onlinecite{Mitra03a}].
In total electron yield detection mode the edge is identical to a
pure CeO$_2$ edge, i.e. cerium in a Ce$^{4+}$ state.
However, when increasing the information depth by switching to bulk
sensitive FY detection,the edge changes drastically.
The FY signal contains contributions from Ce$^{4+}$ and Ce$^{3+}$.
Note that thermodynamically the reducing power of cerium is not
sufficient for the Mn$^{3+}$ - Mn$^{2+}$ transition.
The same trends are seen in the FY spectra of the Mn and O edges.
Manganese reference data were taken from
Ref.~[\onlinecite{Gilbert03}].
In case of the Mn edge [Fig.~\ref{XAS-Ce-Mn}(b)] a decrease of the
Mn$^{2+}$ related feature at 640\,eV is visible in the FY data, and
the edges are broadened towards higher energies than in TEY mode.
This indicates an increased amount of Mn$^{3+}$ (642\,eV) and
Mn$^{4+}$ (644\,eV) species within the film as compared to the near
surface region.
Finally, at the O K edge (not shown here) the relative prepeak
intensity at 530\,eV increases with growing information depth from
TEY to FY mode.
As this feature is proportional to the hole concentration, this
finding further emphasizes the point that the bulk is more oxidized
than the surface and that the majority charge carriers are indeed
holes.

\section{Conclusions}
\label{Sec:Conclusions}

We investigated La$_{0.7}$Ce$_{0.3}$MnO$_3$ thin films of variable
thickness, grown epitaxially at different oxygen pressure $p_{O_2}$
and subjected to different oxygen annealing procedures.
We find that thin film growth at low deposition pressure favors phase
separation via the formation of CeO$_2$ inclusions.
For higher deposition pressure, still CeO$_2$ nanoclusters are found,
as shown by transmission electron microscopy, even for those films
which appear to be single phase in x-ray diffraction analysis.
Combining electric transport, magnetization and Hall measurements
with x-ray photoemission and absorption spectroscopy we obtain a
consistent picture in the sense that the appearance of a
metal-to-insulator transition in electric transport measurements is
always associated with hole doping and the presence of a mixed system
of Mn$^{2+}$, Mn$^{3+}$ and Mn$^{4+}$, despite finding Ce$^{4+}$ as a
sign of electron doping.
The hole-doped behavior of our films may be explained by the presence
of cation vacancies (due to CeO$_2$ clustering), which can be
occupied by excess oxygen that shifts the valences from Mn$^{2+}$ to
Mn$^{3+}$ or Mn$^{4+}$.
In particular, oxidation states are well reproduced in the x-ray
absorption spectra and fit to the transport properties.
Upon oxidizing the samples, the system goes towards Mn$^{3+}$ /
Mn$^{4+}$ as expected, while reducing the films forms more Mn$^{2+}$
species.
In particular for less oxidized films, we find a reduced layer at the
surface with a more oxidized bulk underneath.
This explains some of the peculiarities of this system, namely the
discrepancy between finding Mn$^{2+}$ and Ce$^{4+}$ and still having
holes as majority carriers.
Furthermore, this demonstrates that one has to be very careful in
relating surface sensitive spectroscopy data to bulk sensitive
transport and magnetization data.

\begin{acknowledgments}
We gratefully acknowledge Kathrin D\"{o}rr for helpful discussions and
Matthias Althammer and Sebastian G\"{o}nnenwein for their support with
the Hall measurements.
Furthermore, we acknowledge  the  ANKA  Angstroemquelle  Karlsruhe
for the provision of beamtime and we would like to thank P.~Nagel,
M.~Merz and S.~Schuppler for the skillful technical assistance using
beamline WERA and for valuable discussions.
This work was funded by the Deutsche Forschungsgemeinschaft (project
no.~KO 1303/8-1) and by the European Union under the Framework 6
program for an Integrated Infrastructure Initiative, ref.~026019
ESTEEM.
S.~B. thanks the Fund for Scientific Research -- Flanders.
\end{acknowledgments}


\bibliography{References}

\begin{thebibliography}{36}
\expandafter\ifx\csname natexlab\endcsname\relax\def\natexlab#1{#1}\fi
\expandafter\ifx\csname bibnamefont\endcsname\relax
  \def\bibnamefont#1{#1}\fi
\expandafter\ifx\csname bibfnamefont\endcsname\relax
  \def\bibfnamefont#1{#1}\fi
\expandafter\ifx\csname citenamefont\endcsname\relax
  \def\citenamefont#1{#1}\fi
\expandafter\ifx\csname url\endcsname\relax
  \def\url#1{\texttt{#1}}\fi
\expandafter\ifx\csname urlprefix\endcsname\relax\def\urlprefix{URL }\fi
\providecommand{\bibinfo}[2]{#2}
\providecommand{\eprint}[2][]{\url{#2}}

\bibitem[{\citenamefont{Imada et~al.}(1998)\citenamefont{Imada, Fujimori, and
  Tokura}}]{Imada98}
\bibinfo{author}{\bibfnamefont{M.}~\bibnamefont{Imada}},
  \bibinfo{author}{\bibfnamefont{A.}~\bibnamefont{Fujimori}}, \bibnamefont{and}
  \bibinfo{author}{\bibfnamefont{Y.}~\bibnamefont{Tokura}},
  \bibinfo{journal}{Rev. Mod. Phys.} \textbf{\bibinfo{volume}{70}},
  \bibinfo{pages}{1039} (\bibinfo{year}{1998}).

\bibitem[{\citenamefont{Coey et~al.}(1999)\citenamefont{Coey, Viret, and {von
  Molnar}}}]{Coey99}
\bibinfo{author}{\bibfnamefont{J.~M.~D.} \bibnamefont{Coey}},
  \bibinfo{author}{\bibfnamefont{M.}~\bibnamefont{Viret}}, \bibnamefont{and}
  \bibinfo{author}{\bibfnamefont{S.}~\bibnamefont{{von Molnar}}},
  \bibinfo{journal}{Adv. Phys.} \textbf{\bibinfo{volume}{48}},
  \bibinfo{pages}{167} (\bibinfo{year}{1999}).

\bibitem[{\citenamefont{Salamon and Jaime}(2001)}]{Salamon01}
\bibinfo{author}{\bibfnamefont{M.~B.} \bibnamefont{Salamon}} \bibnamefont{and}
  \bibinfo{author}{\bibfnamefont{M.}~\bibnamefont{Jaime}},
  \bibinfo{journal}{Rev. Mod. Phys.} \textbf{\bibinfo{volume}{73}},
  \bibinfo{pages}{583} (\bibinfo{year}{2001}).

\bibitem[{\citenamefont{Millis}(1998)}]{Millis98}
\bibinfo{author}{\bibfnamefont{A.~J.} \bibnamefont{Millis}},
  \bibinfo{journal}{Nature} \textbf{\bibinfo{volume}{392}},
  \bibinfo{pages}{147} (\bibinfo{year}{1998}).

\bibitem[{\citenamefont{Millis et~al.}(1995)\citenamefont{Millis, Littlewood,
  and Shraiman}}]{Millis95}
\bibinfo{author}{\bibfnamefont{A.~J.} \bibnamefont{Millis}},
  \bibinfo{author}{\bibfnamefont{P.~B.} \bibnamefont{Littlewood}},
  \bibnamefont{and} \bibinfo{author}{\bibfnamefont{B.~I.}
  \bibnamefont{Shraiman}}, \bibinfo{journal}{Phys. Rev. Lett.}
  \textbf{\bibinfo{volume}{74}}, \bibinfo{pages}{5144} (\bibinfo{year}{1995}).

\bibitem[{\citenamefont{Zener}(1951)}]{Zener51}
\bibinfo{author}{\bibfnamefont{C.}~\bibnamefont{Zener}},
  \bibinfo{journal}{Phys. Rev.} \textbf{\bibinfo{volume}{81}},
  \bibinfo{pages}{440} (\bibinfo{year}{1951}).

\bibitem[{\citenamefont{Anderson and Hasegawa}(1955)}]{Anderson55}
\bibinfo{author}{\bibfnamefont{P.~W.} \bibnamefont{Anderson}} \bibnamefont{and}
  \bibinfo{author}{\bibfnamefont{H.}~\bibnamefont{Hasegawa}},
  \bibinfo{journal}{Phys. Rev.} \textbf{\bibinfo{volume}{100}},
  \bibinfo{pages}{675} (\bibinfo{year}{1955}).

\bibitem[{\citenamefont{Jonker and {Van Santen}}(1950)}]{Jonker50}
\bibinfo{author}{\bibfnamefont{G.~H.} \bibnamefont{Jonker}} \bibnamefont{and}
  \bibinfo{author}{\bibfnamefont{J.~H.} \bibnamefont{{Van Santen}}},
  \bibinfo{journal}{Physica} \textbf{\bibinfo{volume}{16}},
  \bibinfo{pages}{337} (\bibinfo{year}{1950}).

\bibitem[{\citenamefont{Mandal and Das}(1997)}]{Mandal97}
\bibinfo{author}{\bibfnamefont{P.}~\bibnamefont{Mandal}} \bibnamefont{and}
  \bibinfo{author}{\bibfnamefont{S.}~\bibnamefont{Das}},
  \bibinfo{journal}{Phys. Rev. B} \textbf{\bibinfo{volume}{56}},
  \bibinfo{pages}{15073} (\bibinfo{year}{1997}).

\bibitem[{\citenamefont{Gebhardt et~al.}(1999)\citenamefont{Gebhardt, Roy, and
  Ali}}]{Gebhardt99}
\bibinfo{author}{\bibfnamefont{J.~R.} \bibnamefont{Gebhardt}},
  \bibinfo{author}{\bibfnamefont{S.}~\bibnamefont{Roy}}, \bibnamefont{and}
  \bibinfo{author}{\bibfnamefont{N.}~\bibnamefont{Ali}}, \bibinfo{journal}{J.
  Appl. Phys.} \textbf{\bibinfo{volume}{85}}, \bibinfo{pages}{5390}
  (\bibinfo{year}{1999}).

\bibitem[{\citenamefont{Ganguly et~al.}(2000)\citenamefont{Ganguly,
  Gopalakrishnan, and Yakhmi}}]{Ganguly00}
\bibinfo{author}{\bibfnamefont{R.}~\bibnamefont{Ganguly}},
  \bibinfo{author}{\bibfnamefont{I.~K.} \bibnamefont{Gopalakrishnan}},
  \bibnamefont{and} \bibinfo{author}{\bibfnamefont{J.~V.}
  \bibnamefont{Yakhmi}}, \bibinfo{journal}{J. Phys.: Condens. Matter}
  \textbf{\bibinfo{volume}{12}}, \bibinfo{pages}{L719} (\bibinfo{year}{2000}).

\bibitem[{\citenamefont{Li et~al.}(1999)\citenamefont{Li, Morrish, and
  Jiang}}]{Li99a}
\bibinfo{author}{\bibfnamefont{Z.~W.} \bibnamefont{Li}},
  \bibinfo{author}{\bibfnamefont{A.~H.} \bibnamefont{Morrish}},
  \bibnamefont{and} \bibinfo{author}{\bibfnamefont{J.~Z.} \bibnamefont{Jiang}},
  \bibinfo{journal}{Phys. Rev. B} \textbf{\bibinfo{volume}{60}},
  \bibinfo{pages}{10284} (\bibinfo{year}{1999}).

\bibitem[{\citenamefont{Tan et~al.}(2003)\citenamefont{Tan, Dai, Duan, Zhou,
  Lu, and Chen}}]{Tan03a}
\bibinfo{author}{\bibfnamefont{G.~T.} \bibnamefont{Tan}},
  \bibinfo{author}{\bibfnamefont{S.}~\bibnamefont{Dai}},
  \bibinfo{author}{\bibfnamefont{P.}~\bibnamefont{Duan}},
  \bibinfo{author}{\bibfnamefont{Y.~L.} \bibnamefont{Zhou}},
  \bibinfo{author}{\bibfnamefont{H.~B.} \bibnamefont{Lu}}, \bibnamefont{and}
  \bibinfo{author}{\bibfnamefont{Z.~H.} \bibnamefont{Chen}},
  \bibinfo{journal}{Phys. Rev. B} \textbf{\bibinfo{volume}{68}},
  \bibinfo{pages}{014426} (\bibinfo{year}{2003}).

\bibitem[{\citenamefont{Mitra et~al.}(2003)\citenamefont{Mitra, Hu,
  Raychaudhuri, Wirth, Csiszar, Hsieh, Lin, Chen, and Tjeng}}]{Mitra03a}
\bibinfo{author}{\bibfnamefont{C.}~\bibnamefont{Mitra}},
  \bibinfo{author}{\bibfnamefont{Z.}~\bibnamefont{Hu}},
  \bibinfo{author}{\bibfnamefont{P.}~\bibnamefont{Raychaudhuri}},
  \bibinfo{author}{\bibfnamefont{S.}~\bibnamefont{Wirth}},
  \bibinfo{author}{\bibfnamefont{S.~I.} \bibnamefont{Csiszar}},
  \bibinfo{author}{\bibfnamefont{H.~H.} \bibnamefont{Hsieh}},
  \bibinfo{author}{\bibfnamefont{H.-J.} \bibnamefont{Lin}},
  \bibinfo{author}{\bibfnamefont{C.~T.} \bibnamefont{Chen}}, \bibnamefont{and}
  \bibinfo{author}{\bibfnamefont{L.~H.} \bibnamefont{Tjeng}},
  \bibinfo{journal}{Phys. Rev. B} \textbf{\bibinfo{volume}{67}},
  \bibinfo{pages}{092404} (\bibinfo{year}{2003}).

\bibitem[{\citenamefont{Philip and Kutty}(1999)}]{Philip99}
\bibinfo{author}{\bibfnamefont{J.}~\bibnamefont{Philip}} \bibnamefont{and}
  \bibinfo{author}{\bibfnamefont{T.~R.~N.} \bibnamefont{Kutty}},
  \bibinfo{journal}{J. Phys.: Condens. Matter} \textbf{\bibinfo{volume}{11}},
  \bibinfo{pages}{8537} (\bibinfo{year}{1999}).

\bibitem[{\citenamefont{Mitra et~al.}(2001{\natexlab{a}})\citenamefont{Mitra,
  Raychaudhuri, K{\"o}bernik, D{\"o}rr, M{\"u}ller, and Schultz}}]{Mitra01a}
\bibinfo{author}{\bibfnamefont{C.}~\bibnamefont{Mitra}},
  \bibinfo{author}{\bibfnamefont{P.}~\bibnamefont{Raychaudhuri}},
  \bibinfo{author}{\bibfnamefont{G.}~\bibnamefont{K{\"o}bernik}},
  \bibinfo{author}{\bibfnamefont{K.}~\bibnamefont{D{\"o}rr}},
  \bibinfo{author}{\bibfnamefont{K.-H.} \bibnamefont{M{\"u}ller}},
  \bibnamefont{and} \bibinfo{author}{\bibfnamefont{L.}~\bibnamefont{Schultz}},
  \bibinfo{journal}{Appl. Phys. Lett.} \textbf{\bibinfo{volume}{79}},
  \bibinfo{pages}{2408} (\bibinfo{year}{2001}{\natexlab{a}}).

\bibitem[{\citenamefont{Raychaudhuri et~al.}(1999)\citenamefont{Raychaudhuri,
  Mukherjee, Nigam, John, Vaisnav, and Pinto}}]{Raychaudhuri99}
\bibinfo{author}{\bibfnamefont{P.}~\bibnamefont{Raychaudhuri}},
  \bibinfo{author}{\bibfnamefont{S.}~\bibnamefont{Mukherjee}},
  \bibinfo{author}{\bibfnamefont{A.~K.} \bibnamefont{Nigam}},
  \bibinfo{author}{\bibfnamefont{J.}~\bibnamefont{John}},
  \bibinfo{author}{\bibfnamefont{U.~D.} \bibnamefont{Vaisnav}},
  \bibnamefont{and} \bibinfo{author}{\bibfnamefont{R.}~\bibnamefont{Pinto}},
  \bibinfo{journal}{J. Appl. Phys.} \textbf{\bibinfo{volume}{86}},
  \bibinfo{pages}{5718} (\bibinfo{year}{1999}).

\bibitem[{\citenamefont{Han et~al.}(2004)\citenamefont{Han, Lee, Kim, Mitra,
  Jeong, Kim, Kim, Min, Kim, Wi et~al.}}]{Han04}
\bibinfo{author}{\bibfnamefont{S.~W.} \bibnamefont{Han}},
  \bibinfo{author}{\bibfnamefont{J.~D.} \bibnamefont{Lee}},
  \bibinfo{author}{\bibfnamefont{K.~H.} \bibnamefont{Kim}},
  \bibinfo{author}{\bibfnamefont{C.}~\bibnamefont{Mitra}},
  \bibinfo{author}{\bibfnamefont{J.~I.} \bibnamefont{Jeong}},
  \bibinfo{author}{\bibfnamefont{K.~J.} \bibnamefont{Kim}},
  \bibinfo{author}{\bibfnamefont{B.~S.} \bibnamefont{Kim}},
  \bibinfo{author}{\bibfnamefont{B.~I.} \bibnamefont{Min}},
  \bibinfo{author}{\bibfnamefont{J.~H.} \bibnamefont{Kim}},
  \bibinfo{author}{\bibfnamefont{S.~C.} \bibnamefont{Wi}},
  \bibnamefont{et~al.}, \bibinfo{journal}{Phys. stat. sol.}
  \textbf{\bibinfo{volume}{241}}, \bibinfo{pages}{1577} (\bibinfo{year}{2004}).

\bibitem[{\citenamefont{Wang et~al.}(2006)\citenamefont{Wang, Sun, Zhang, Liu,
  Shen, Tian, and Li}}]{Wang06}
\bibinfo{author}{\bibfnamefont{D.~J.} \bibnamefont{Wang}},
  \bibinfo{author}{\bibfnamefont{J.~R.} \bibnamefont{Sun}},
  \bibinfo{author}{\bibfnamefont{S.~Y.} \bibnamefont{Zhang}},
  \bibinfo{author}{\bibfnamefont{G.~J.} \bibnamefont{Liu}},
  \bibinfo{author}{\bibfnamefont{B.~G.} \bibnamefont{Shen}},
  \bibinfo{author}{\bibfnamefont{H.~F.} \bibnamefont{Tian}}, \bibnamefont{and}
  \bibinfo{author}{\bibfnamefont{J.~Q.} \bibnamefont{Li}},
  \bibinfo{journal}{Phys. Rev. B} \textbf{\bibinfo{volume}{73}},
  \bibinfo{pages}{144403} (\bibinfo{year}{2006}).

\bibitem[{\citenamefont{Zhao et~al.}(2000)\citenamefont{Zhao, Srivastava,
  Fournier, Smolyaninova, Rajeswari, Wu, Li, Greene, and Venkatesan}}]{Zhao00}
\bibinfo{author}{\bibfnamefont{Y.}~\bibnamefont{Zhao}},
  \bibinfo{author}{\bibfnamefont{R.}~\bibnamefont{Srivastava}},
  \bibinfo{author}{\bibfnamefont{P.}~\bibnamefont{Fournier}},
  \bibinfo{author}{\bibfnamefont{V.}~\bibnamefont{Smolyaninova}},
  \bibinfo{author}{\bibfnamefont{M.}~\bibnamefont{Rajeswari}},
  \bibinfo{author}{\bibfnamefont{T.}~\bibnamefont{Wu}},
  \bibinfo{author}{\bibfnamefont{Z.}~\bibnamefont{Li}},
  \bibinfo{author}{\bibfnamefont{R.}~\bibnamefont{Greene}}, \bibnamefont{and}
  \bibinfo{author}{\bibfnamefont{T.}~\bibnamefont{Venkatesan}},
  \bibinfo{journal}{J. Magn. Magn. Mater.} \textbf{\bibinfo{volume}{220}},
  \bibinfo{pages}{161} (\bibinfo{year}{2000}).

\bibitem[{\citenamefont{Yanagida et~al.}(2004)\citenamefont{Yanagida, Kanki,
  Vilquin, Tanaka, and Kawai}}]{Yanagida04}
\bibinfo{author}{\bibfnamefont{T.}~\bibnamefont{Yanagida}},
  \bibinfo{author}{\bibfnamefont{T.}~\bibnamefont{Kanki}},
  \bibinfo{author}{\bibfnamefont{B.}~\bibnamefont{Vilquin}},
  \bibinfo{author}{\bibfnamefont{H.}~\bibnamefont{Tanaka}}, \bibnamefont{and}
  \bibinfo{author}{\bibfnamefont{T.}~\bibnamefont{Kawai}},
  \bibinfo{journal}{Phys. Rev. B} \textbf{\bibinfo{volume}{70}},
  \bibinfo{pages}{184437} (\bibinfo{year}{2004}).

\bibitem[{\citenamefont{Yanagida et~al.}(2005)\citenamefont{Yanagida, Kanki,
  and Vilquin}}]{Yanagida05}
\bibinfo{author}{\bibfnamefont{T.}~\bibnamefont{Yanagida}},
  \bibinfo{author}{\bibfnamefont{T.}~\bibnamefont{Kanki}}, \bibnamefont{and}
  \bibinfo{author}{\bibfnamefont{B.}~\bibnamefont{Vilquin}},
  \bibinfo{journal}{J. App. Phys.} \textbf{\bibinfo{volume}{97}},
  \bibinfo{pages}{33905} (\bibinfo{year}{2005}).

\bibitem[{\citenamefont{Mitra et~al.}(2001{\natexlab{b}})\citenamefont{Mitra,
  Raychaudhuri, John, Dhar, Nigam, and Pinto}}]{Mitra01}
\bibinfo{author}{\bibfnamefont{C.}~\bibnamefont{Mitra}},
  \bibinfo{author}{\bibfnamefont{P.}~\bibnamefont{Raychaudhuri}},
  \bibinfo{author}{\bibfnamefont{J.}~\bibnamefont{John}},
  \bibinfo{author}{\bibfnamefont{S.~K.} \bibnamefont{Dhar}},
  \bibinfo{author}{\bibfnamefont{A.~K.} \bibnamefont{Nigam}}, \bibnamefont{and}
  \bibinfo{author}{\bibfnamefont{R.}~\bibnamefont{Pinto}}, \bibinfo{journal}{J.
  App. Phys.} \textbf{\bibinfo{volume}{89}}, \bibinfo{pages}{524}
  (\bibinfo{year}{2001}{\natexlab{b}}).

\bibitem[{\citenamefont{Chang et~al.}(2004)\citenamefont{Chang, C.~C.~Hsieh,
  Wu, Uen, Gou, Hsu, and Lin}}]{Chang04}
\bibinfo{author}{\bibfnamefont{W.~J.} \bibnamefont{Chang}},
  \bibinfo{author}{\bibfnamefont{J.~Y.~J.} \bibnamefont{C.~C.~Hsieh}},
  \bibinfo{author}{\bibfnamefont{K.~H.} \bibnamefont{Wu}},
  \bibinfo{author}{\bibfnamefont{T.~M.} \bibnamefont{Uen}},
  \bibinfo{author}{\bibfnamefont{Y.~S.} \bibnamefont{Gou}},
  \bibinfo{author}{\bibfnamefont{C.~H.} \bibnamefont{Hsu}}, \bibnamefont{and}
  \bibinfo{author}{\bibfnamefont{J.-Y.} \bibnamefont{Lin}},
  \bibinfo{journal}{J. Appl. Phys.} \textbf{\bibinfo{volume}{96}},
  \bibinfo{pages}{4357} (\bibinfo{year}{2004}).

\bibitem[{\citenamefont{T{\"o}pfer and Goodenough}(1997)}]{Toepfer97}
\bibinfo{author}{\bibfnamefont{J.}~\bibnamefont{T{\"o}pfer}} \bibnamefont{and}
  \bibinfo{author}{\bibfnamefont{J.~B.} \bibnamefont{Goodenough}},
  \bibinfo{journal}{J. Solid State Chem.} \textbf{\bibinfo{volume}{130}},
  \bibinfo{pages}{117} (\bibinfo{year}{1997}).

\bibitem[{\citenamefont{Murugavel et~al.}(2003)\citenamefont{Murugavel, Lee,
  Yoon, Noh, Chung, Heu, and Yoon}}]{Murugavel03}
\bibinfo{author}{\bibfnamefont{P.}~\bibnamefont{Murugavel}},
  \bibinfo{author}{\bibfnamefont{J.~H.} \bibnamefont{Lee}},
  \bibinfo{author}{\bibfnamefont{J.-G.} \bibnamefont{Yoon}},
  \bibinfo{author}{\bibfnamefont{T.~W.} \bibnamefont{Noh}},
  \bibinfo{author}{\bibfnamefont{J.-S.} \bibnamefont{Chung}},
  \bibinfo{author}{\bibfnamefont{M.}~\bibnamefont{Heu}}, \bibnamefont{and}
  \bibinfo{author}{\bibfnamefont{S.}~\bibnamefont{Yoon}},
  \bibinfo{journal}{Appl. Phys. Lett.} \textbf{\bibinfo{volume}{82}},
  \bibinfo{pages}{1908} (\bibinfo{year}{2003}).

\bibitem[{\citenamefont{Huijbregtse et~al.}(2001)\citenamefont{Huijbregtse,
  Rector, and Dam}}]{Huijbregtse01}
\bibinfo{author}{\bibfnamefont{J.~M.} \bibnamefont{Huijbregtse}},
  \bibinfo{author}{\bibfnamefont{J.~H.} \bibnamefont{Rector}},
  \bibnamefont{and} \bibinfo{author}{\bibfnamefont{B.}~\bibnamefont{Dam}},
  \bibinfo{journal}{Physica C} \textbf{\bibinfo{volume}{351}},
  \bibinfo{pages}{183} (\bibinfo{year}{2001}).

\bibitem[{\citenamefont{Koster et~al.}(1998)\citenamefont{Koster, Kropman,
  Rijnders, Blank, and Rogalla}}]{Koster98}
\bibinfo{author}{\bibfnamefont{G.}~\bibnamefont{Koster}},
  \bibinfo{author}{\bibfnamefont{B.~L.} \bibnamefont{Kropman}},
  \bibinfo{author}{\bibfnamefont{G.~J. H.~M.} \bibnamefont{Rijnders}},
  \bibinfo{author}{\bibfnamefont{D.~H.~A.} \bibnamefont{Blank}},
  \bibnamefont{and} \bibinfo{author}{\bibfnamefont{H.}~\bibnamefont{Rogalla}},
  \bibinfo{journal}{Appl. Phys. Lett.} \textbf{\bibinfo{volume}{73}},
  \bibinfo{pages}{2920} (\bibinfo{year}{1998}).

\bibitem[{\citenamefont{Zhang and Zhang}(2003)}]{Zhang03}
\bibinfo{author}{\bibfnamefont{Q.}~\bibnamefont{Zhang}} \bibnamefont{and}
  \bibinfo{author}{\bibfnamefont{W.}~\bibnamefont{Zhang}},
  \bibinfo{journal}{Phys. Rev. B} \textbf{\bibinfo{volume}{68}},
  \bibinfo{pages}{134449} (\bibinfo{year}{2003}).

\bibitem[{\citenamefont{Karplus and Luttinger}(1954)}]{Karplus54}
\bibinfo{author}{\bibfnamefont{R.}~\bibnamefont{Karplus}} \bibnamefont{and}
  \bibinfo{author}{\bibfnamefont{J.~M.} \bibnamefont{Luttinger}},
  \bibinfo{journal}{Phys. Rev.} \textbf{\bibinfo{volume}{95}},
  \bibinfo{pages}{1154} (\bibinfo{year}{1954}).

\bibitem[{\citenamefont{Abbate et~al.}(1992)\citenamefont{Abbate, {de Groot},
  Guggle, A, Strebel, Lopez, Domke, Kaindl, Sawatzky, Takano
  et~al.}}]{Abbate92}
\bibinfo{author}{\bibfnamefont{M.}~\bibnamefont{Abbate}},
  \bibinfo{author}{\bibfnamefont{F.~M.~F.} \bibnamefont{{de Groot}}},
  \bibinfo{author}{\bibfnamefont{J.~C.} \bibnamefont{Guggle}},
  \bibinfo{author}{\bibfnamefont{F.}~\bibnamefont{A}},
  \bibinfo{author}{\bibfnamefont{O.}~\bibnamefont{Strebel}},
  \bibinfo{author}{\bibfnamefont{F.}~\bibnamefont{Lopez}},
  \bibinfo{author}{\bibfnamefont{M.}~\bibnamefont{Domke}},
  \bibinfo{author}{\bibfnamefont{G.}~\bibnamefont{Kaindl}},
  \bibinfo{author}{\bibfnamefont{G.~A.} \bibnamefont{Sawatzky}},
  \bibinfo{author}{\bibfnamefont{M.}~\bibnamefont{Takano}},
  \bibnamefont{et~al.}, \bibinfo{journal}{Phys. Rev. B}
  \textbf{\bibinfo{volume}{46}}, \bibinfo{pages}{4511} (\bibinfo{year}{1992}).

\bibitem[{\citenamefont{Mannella et~al.}(2005)\citenamefont{Mannella,
  Rosenhahn, Watanabe, Sell, Nambu, Ritchey, Arenholz, Young, Tomioka, and
  Fadley}}]{Manella05}
\bibinfo{author}{\bibfnamefont{N.}~\bibnamefont{Mannella}},
  \bibinfo{author}{\bibfnamefont{A.}~\bibnamefont{Rosenhahn}},
  \bibinfo{author}{\bibfnamefont{M.}~\bibnamefont{Watanabe}},
  \bibinfo{author}{\bibfnamefont{B.}~\bibnamefont{Sell}},
  \bibinfo{author}{\bibfnamefont{A.}~\bibnamefont{Nambu}},
  \bibinfo{author}{\bibfnamefont{S.}~\bibnamefont{Ritchey}},
  \bibinfo{author}{\bibfnamefont{E.}~\bibnamefont{Arenholz}},
  \bibinfo{author}{\bibfnamefont{A.}~\bibnamefont{Young}},
  \bibinfo{author}{\bibfnamefont{Y.}~\bibnamefont{Tomioka}}, \bibnamefont{and}
  \bibinfo{author}{\bibfnamefont{C.~S.} \bibnamefont{Fadley}},
  \bibinfo{journal}{Phys. Rev. B} \textbf{\bibinfo{volume}{71}},
  \bibinfo{pages}{125117} (\bibinfo{year}{2005}).

\bibitem[{\citenamefont{Chang et~al.}(2005)\citenamefont{Chang, Tsai, Jeng,
  Lin, Zhang, Liu, Lee, Chen, Wu, Uen et~al.}}]{Chang05}
\bibinfo{author}{\bibfnamefont{W.~J.} \bibnamefont{Chang}},
  \bibinfo{author}{\bibfnamefont{J.~Y.} \bibnamefont{Tsai}},
  \bibinfo{author}{\bibfnamefont{H.-T.} \bibnamefont{Jeng}},
  \bibinfo{author}{\bibfnamefont{J.-Y.} \bibnamefont{Lin}},
  \bibinfo{author}{\bibfnamefont{K.~Y.-J.} \bibnamefont{Zhang}},
  \bibinfo{author}{\bibfnamefont{H.~L.} \bibnamefont{Liu}},
  \bibinfo{author}{\bibfnamefont{J.~M.} \bibnamefont{Lee}},
  \bibinfo{author}{\bibfnamefont{J.~M.} \bibnamefont{Chen}},
  \bibinfo{author}{\bibfnamefont{K.~H.} \bibnamefont{Wu}},
  \bibinfo{author}{\bibfnamefont{T.~M.} \bibnamefont{Uen}},
  \bibnamefont{et~al.}, \bibinfo{journal}{Phys. Rev. B}
  \textbf{\bibinfo{volume}{72}}, \bibinfo{pages}{132410}
  (\bibinfo{year}{2005}).

\bibitem[{\citenamefont{Nagel et~al.}(2007)\citenamefont{Nagel, Biswas, Nagel,
  Pellegrin, Schuppler, Peisert, and Chass{\'e}}}]{Nagel07}
\bibinfo{author}{\bibfnamefont{M.}~\bibnamefont{Nagel}},
  \bibinfo{author}{\bibfnamefont{I.}~\bibnamefont{Biswas}},
  \bibinfo{author}{\bibfnamefont{P.}~\bibnamefont{Nagel}},
  \bibinfo{author}{\bibfnamefont{E.}~\bibnamefont{Pellegrin}},
  \bibinfo{author}{\bibfnamefont{S.}~\bibnamefont{Schuppler}},
  \bibinfo{author}{\bibfnamefont{H.}~\bibnamefont{Peisert}}, \bibnamefont{and}
  \bibinfo{author}{\bibfnamefont{T.}~\bibnamefont{Chass{\'e}}},
  \bibinfo{journal}{Phys. Rev. B} \textbf{\bibinfo{volume}{75}},
  \bibinfo{pages}{195426} (\bibinfo{year}{2007}).

\bibitem[{\citenamefont{Gilbert et~al.}(2003)\citenamefont{Gilbert, Frazer,
  Belz, Conrad, Nealson, Haskel, Lang, Srajer, and Stasio}}]{Gilbert03}
\bibinfo{author}{\bibfnamefont{B.}~\bibnamefont{Gilbert}},
  \bibinfo{author}{\bibfnamefont{B.~H.} \bibnamefont{Frazer}},
  \bibinfo{author}{\bibfnamefont{A.}~\bibnamefont{Belz}},
  \bibinfo{author}{\bibfnamefont{P.~G.} \bibnamefont{Conrad}},
  \bibinfo{author}{\bibfnamefont{K.~H.} \bibnamefont{Nealson}},
  \bibinfo{author}{\bibfnamefont{D.}~\bibnamefont{Haskel}},
  \bibinfo{author}{\bibfnamefont{J.~C.} \bibnamefont{Lang}},
  \bibinfo{author}{\bibfnamefont{G.}~\bibnamefont{Srajer}}, \bibnamefont{and}
  \bibinfo{author}{\bibfnamefont{G.~D.} \bibnamefont{Stasio}},
  \bibinfo{journal}{J. Phys. Chem. A} \textbf{\bibinfo{volume}{107}},
  \bibinfo{pages}{2839} (\bibinfo{year}{2003}).

\bibitem[{\citenamefont{Tr{\"o}ger et~al.}(1992)\citenamefont{Tr{\"o}ger,
  Arvanitis, Baberschke, Michaelis, Grimm, and Zschech}}]{Troeger92}
\bibinfo{author}{\bibfnamefont{L.}~\bibnamefont{Tr{\"o}ger}},
  \bibinfo{author}{\bibfnamefont{D.}~\bibnamefont{Arvanitis}},
  \bibinfo{author}{\bibfnamefont{K.}~\bibnamefont{Baberschke}},
  \bibinfo{author}{\bibfnamefont{H.}~\bibnamefont{Michaelis}},
  \bibinfo{author}{\bibfnamefont{U.}~\bibnamefont{Grimm}}, \bibnamefont{and}
  \bibinfo{author}{\bibfnamefont{E.}~\bibnamefont{Zschech}},
  \bibinfo{journal}{Phys. Rev. B} \textbf{\bibinfo{volume}{46}},
  \bibinfo{pages}{3283} (\bibinfo{year}{1992}).

\end{thebibliography}

\end{document}

\clearpage
\newpage

\textbf{\Large Figures}

\figtrue \setcounter{figure}{0}

\begin{figure}
\iffig
\centering
\includegraphics[width=0.45\textwidth]{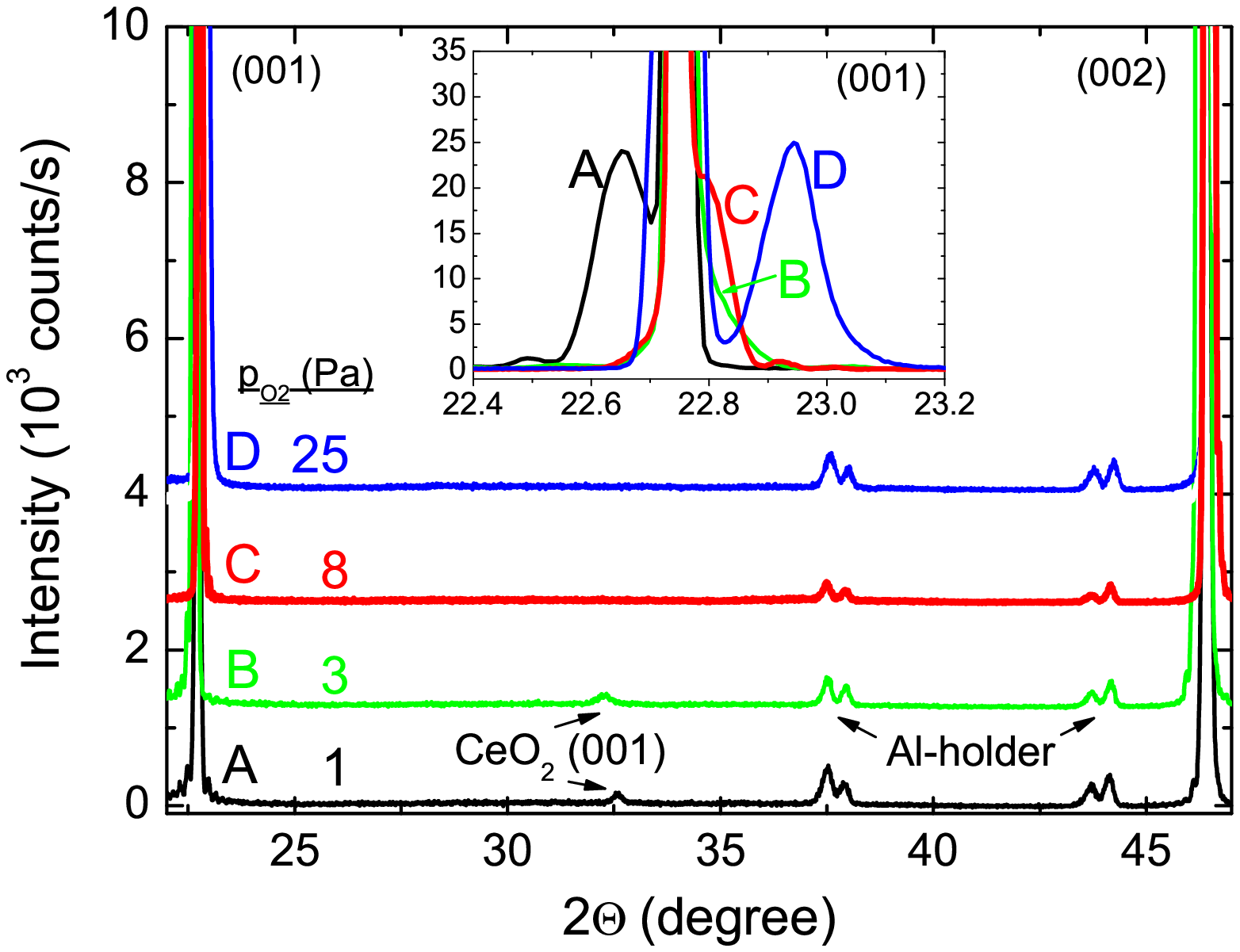}
\fi
\ifnote{\sf\textcolor{red}{[filename: XRDCeO2]\;}}\fi
\caption{\ifnote{\sf\textcolor{red}{[label: XRDCeO2]\;}}\fi
(Color online)
XRD patterns of samples grown under different deposition pressures:
$p_{O_2}=1,\,3,\,8$ and $25\,$Pa for sample A, B, C and D, respectively.
CeO$_2$ can be identified in samples A and B.
XRD scans are offset for clarity.
The inset shows a detailed view around the (001) substrate peak
including the (001) film peaks.}
%
\end{figure}

\begin{figure}
\iffig
\centering
\includegraphics[width=0.45\textwidth]{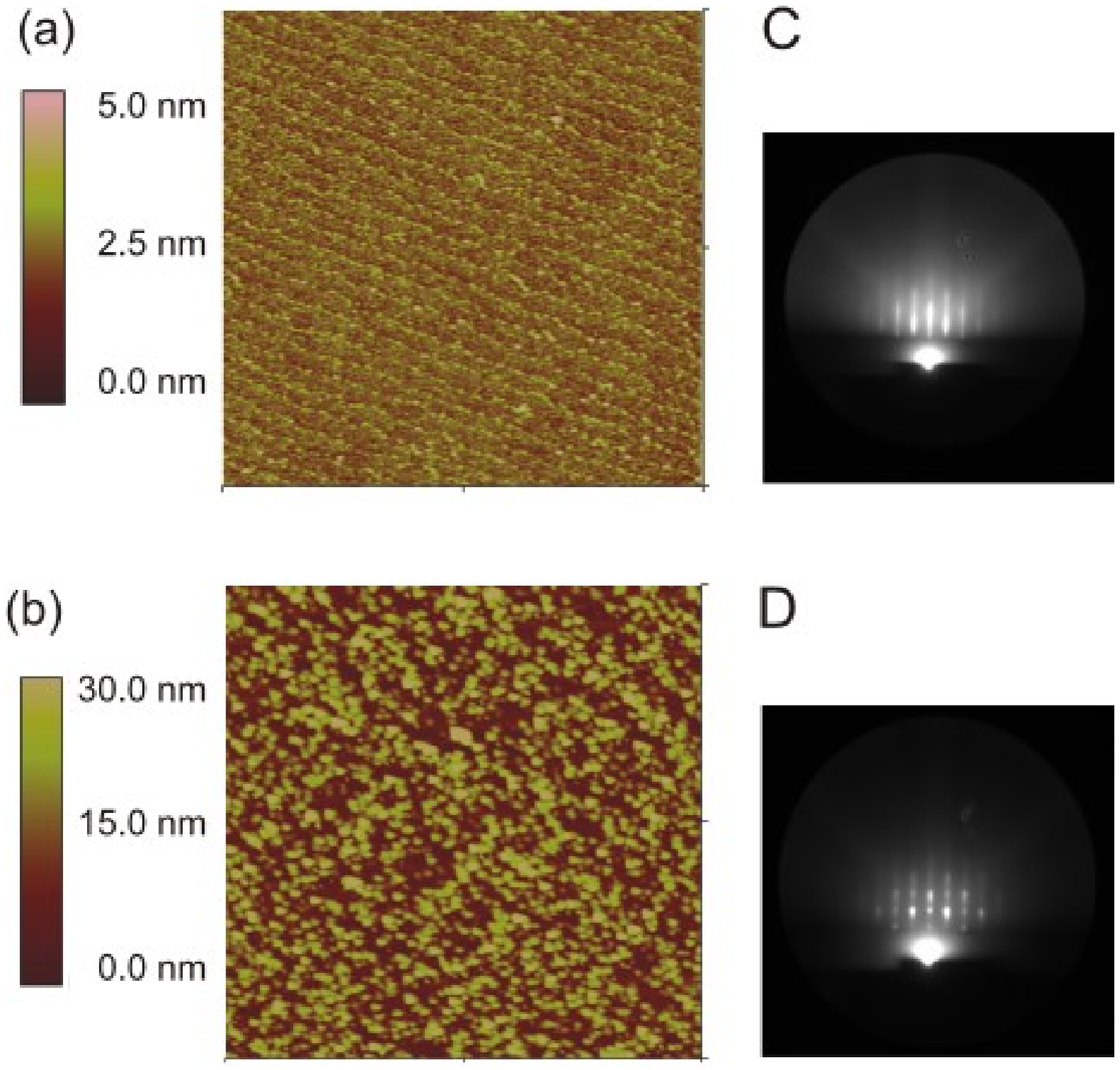}
\fi
\ifnote{\sf\textcolor{red}{[AFM-RHEED]\;}}\fi
\caption{\ifnote{\sf\textcolor{red}{[AFM-RHEED]\;}}\fi
(Color online)
AFM images (left; frame size $5\times 5\,\mu{\rm m}^2$)
and RHEED images (right) of 100\,nm thick films:
(a) sample C, grown at $p_{O_2}=8\,$Pa, and
(b) sample D, grown at $p_{O_2}=25\,$Pa.}
%
\end{figure}

\begin{figure}
\iffig
\includegraphics[width=0.35\textwidth]{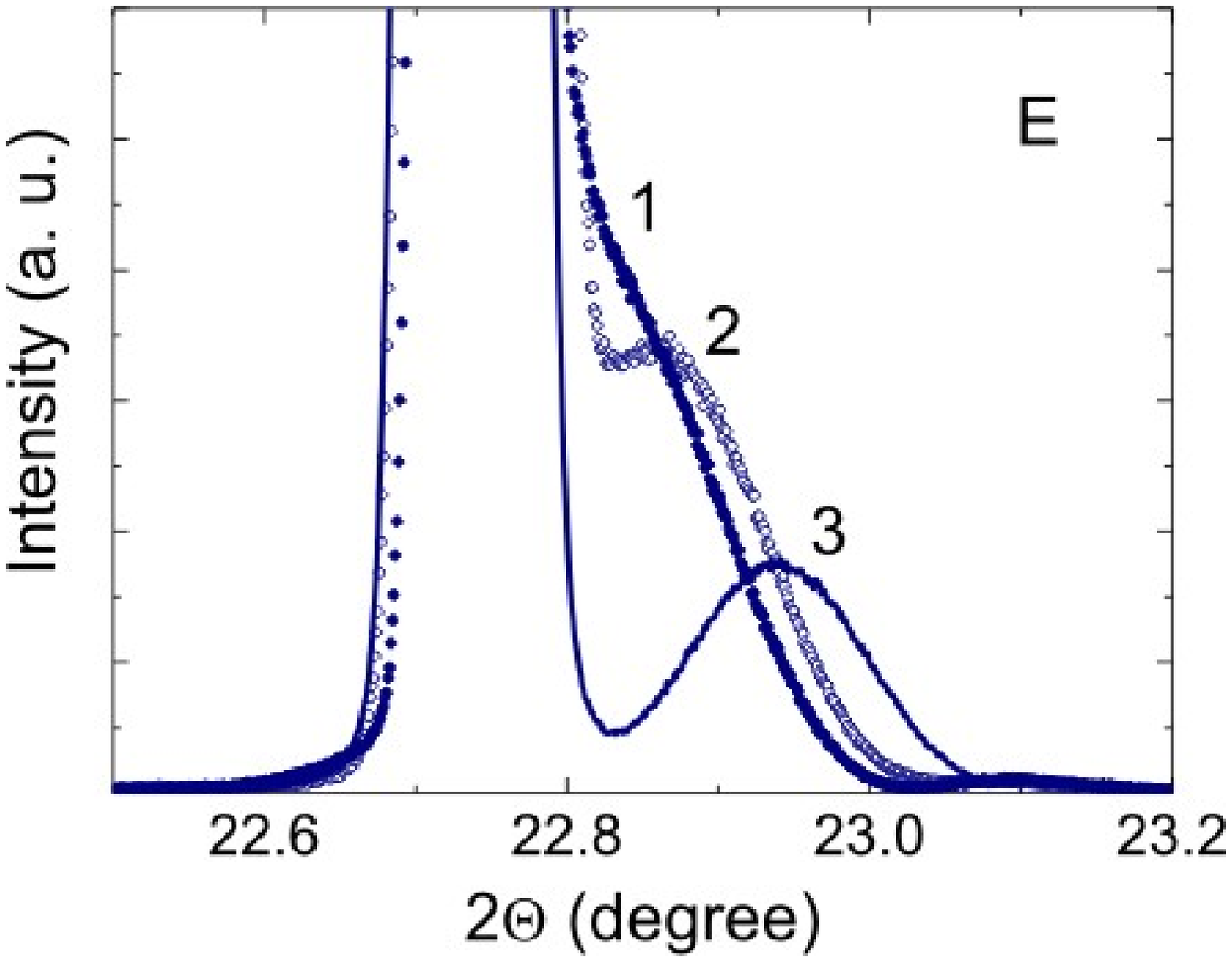}
\fi
\ifnote{\sf\textcolor{red}{[XRDrr33annealing]\;}}\fi
\caption{\ifnote{\sf\textcolor{red}{[annealing]\;}}\fi
(Color online)
XRD pattern at the (001) peak for sample E, showing the evolution of
the $c$-axis with ex-situ annealing steps: after in-situ annealing
(1), first (2) and second (3) ex-situ annealing.}
%
\end{figure}

\begin{figure}
\iffig
\centering
\includegraphics[width=0.45\textwidth]{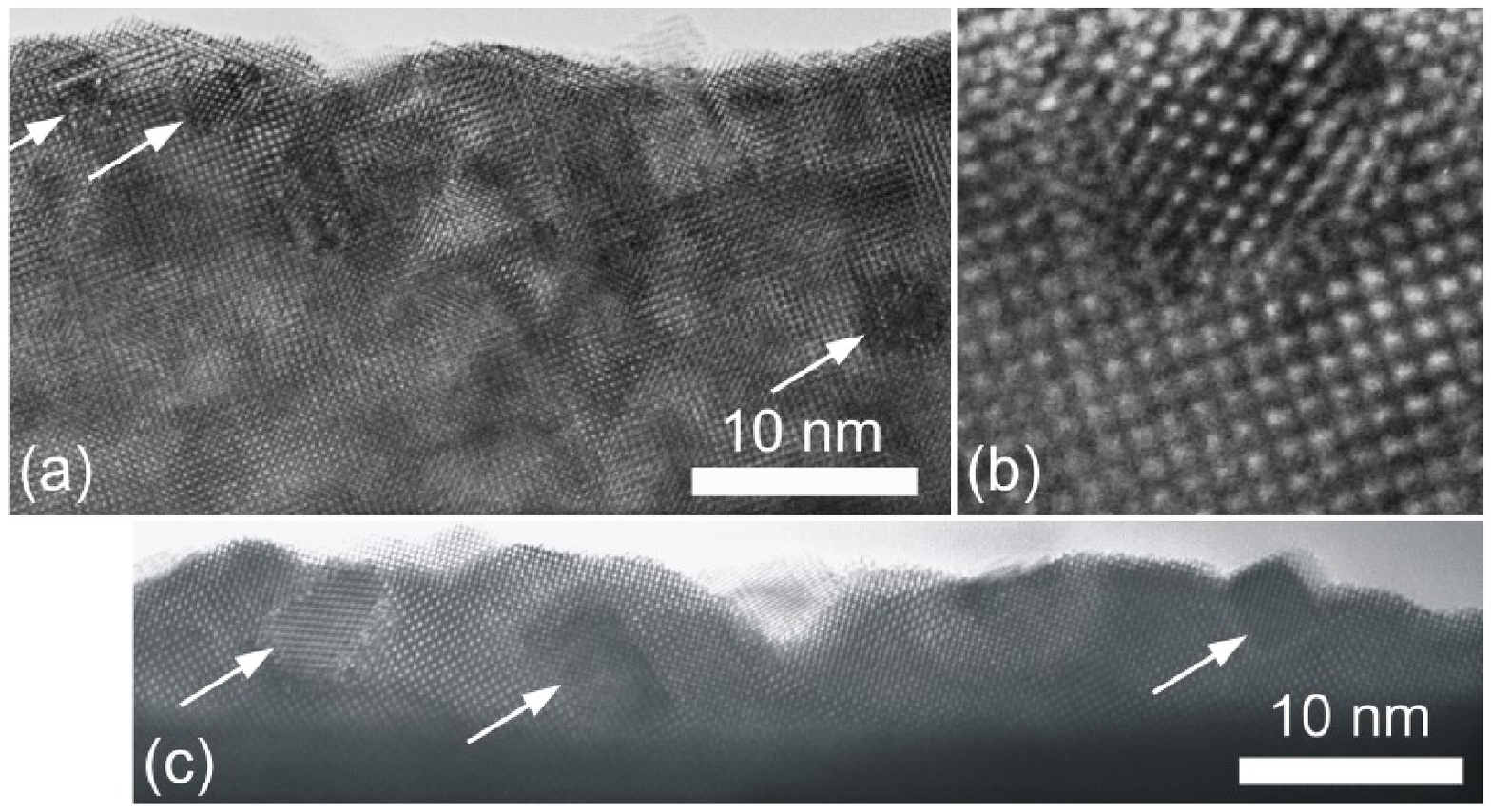}
\fi
\ifnote{\sf\textcolor{red}{[TEM]\;}}\fi
\caption{\ifnote{\sf\textcolor{red}{[TEM]\;}}\fi
(a) Planview HRTEM images of (a) sample E
grown at 8\,Pa O$_2$;
arrows indicate the CeO$_2$ inclusions.
An example of an inclusion [cf.~left arrow in (a)]
is shown in more detail in (b).
(c) sample K grown at 8\,Pa O$_2$.}
%
\end{figure}

\begin{figure}
\iffig
\centering
\includegraphics[width=0.45\textwidth]{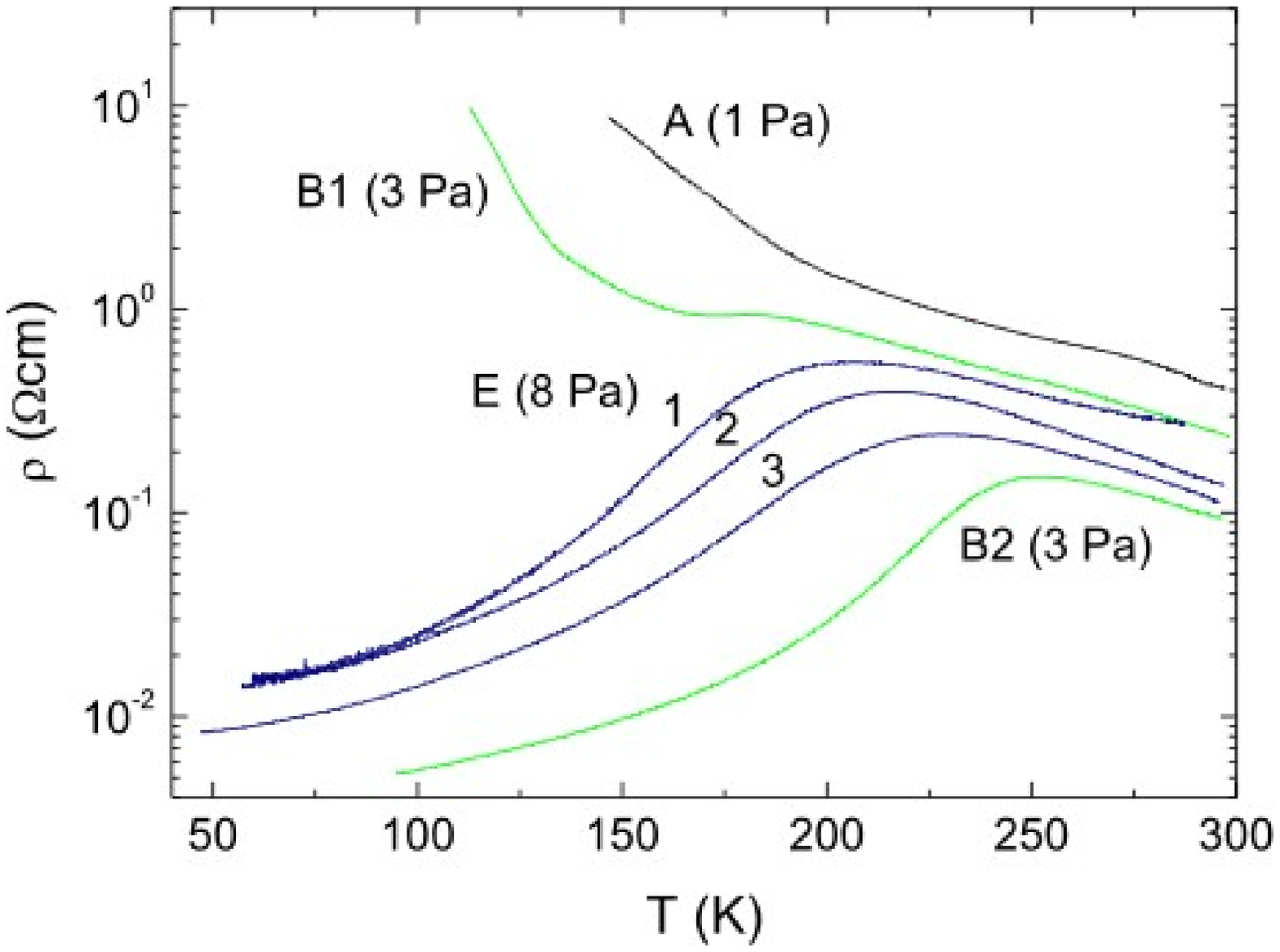}
\fi
\ifnote{\sf\textcolor{red}{[RT]\;}}\fi
\caption{\ifnote{\sf\textcolor{red}{[RT]\;}}\fi
(Color online)
Resistivity vs.~temperature for samples A, B and E (with deposition
pressure $p_{O_2}$ in parenthesis).
The behavior after ex-situ annealing is shown for sample B and E
(B1, E1: without ex-situ annealing;
B2, E2: after $1^{\rm st}$ ex-situ annealing step;
E3: after $2^{\rm nd}$ ex-situ annealing step).}
%
\end{figure}

\begin{figure}
\iffig
\centering
\includegraphics[width=0.35\textwidth]{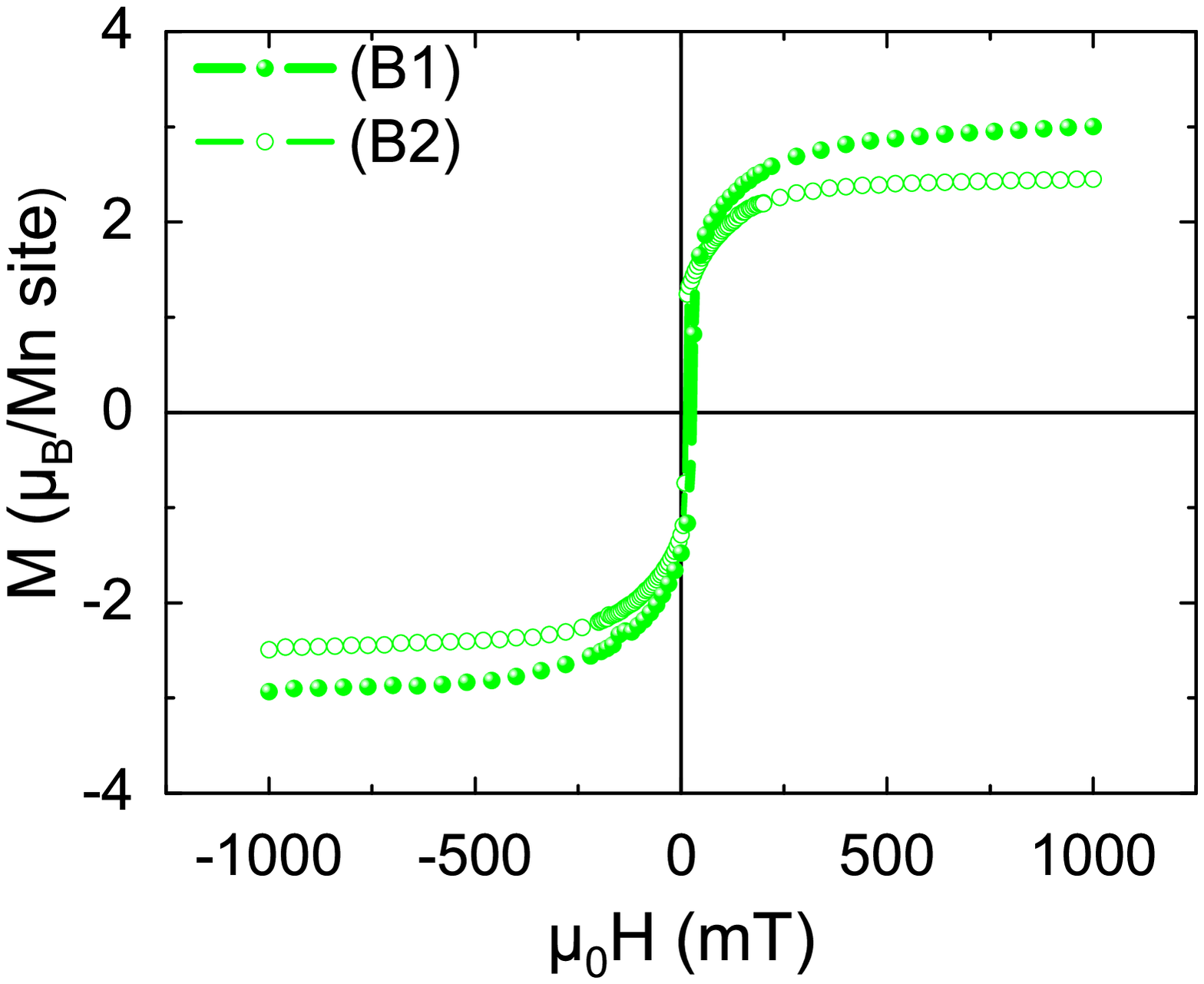}
\fi
\ifnote{\sf\textcolor{red}{[MvsH1]\;}}\fi
\caption{\ifnote{\sf\textcolor{red}{[MvsH1]\;}}\fi
(Color online)
Magnetization vs.~applied magnetic field at $T=20\,$K
for the as-grown ($p_{O_2}=3\,$Pa) sample B (B1)
and after ex-situ annealing (B2).}
%
\end{figure}

\begin{figure}
\iffig
\centering
\includegraphics[width=0.45\textwidth]{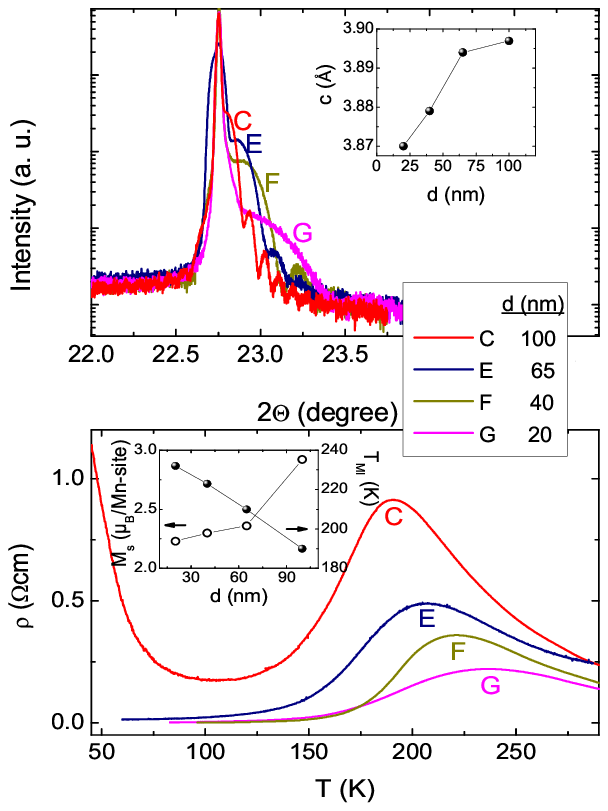}
\fi
\ifnote{\sf\textcolor{red}{[thickness]\;}}\fi
\caption{\ifnote{\sf\textcolor{red}{[thickness]\;}}\fi
(Color online)
Comparison of samples with different thickness $d$, grown under the
same deposition conditions ($p_{O_2}=8\,$Pa; in-situ annealed).
(a) XRD $\Theta - 2\Theta$ scans; the inset shows that the $c$-axis value increases with increasing $d$.
(b) Resistivity vs.~temperature; the inset shows that the transition temperature $T_{MI}$
decreases and the saturation magnetization $M_s$ (from $M(H)$ data; not shown) increases with increasing $d$.}
%
\end{figure}

\begin{figure}
\iffig
\centering
\includegraphics[width=0.35\textwidth]{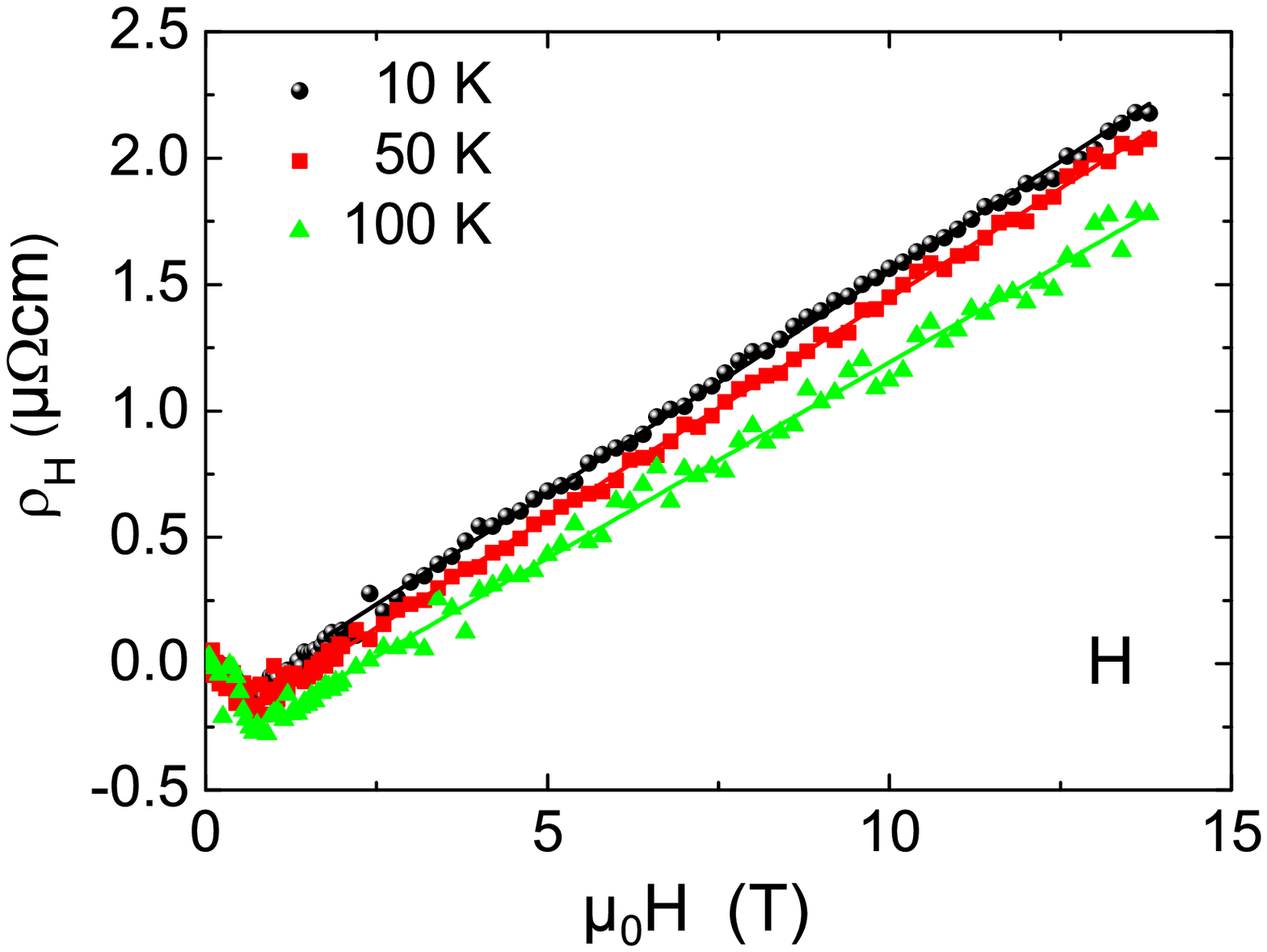}
\fi
\ifnote{\sf\textcolor{red}{[Hall]\;}}\fi
\caption{\ifnote{\sf\textcolor{red}{[Hall]\;}}\fi
(Color online)
Field dependence of the Hall resistivity of sample H.
The positive slope at high magnetic field identifies the majority of
the carriers to be holes.
The solid lines are linear fits to the high-field data.}
%
\end{figure}

\begin{figure}
\iffig
\centering
\includegraphics[width=0.45\textwidth]{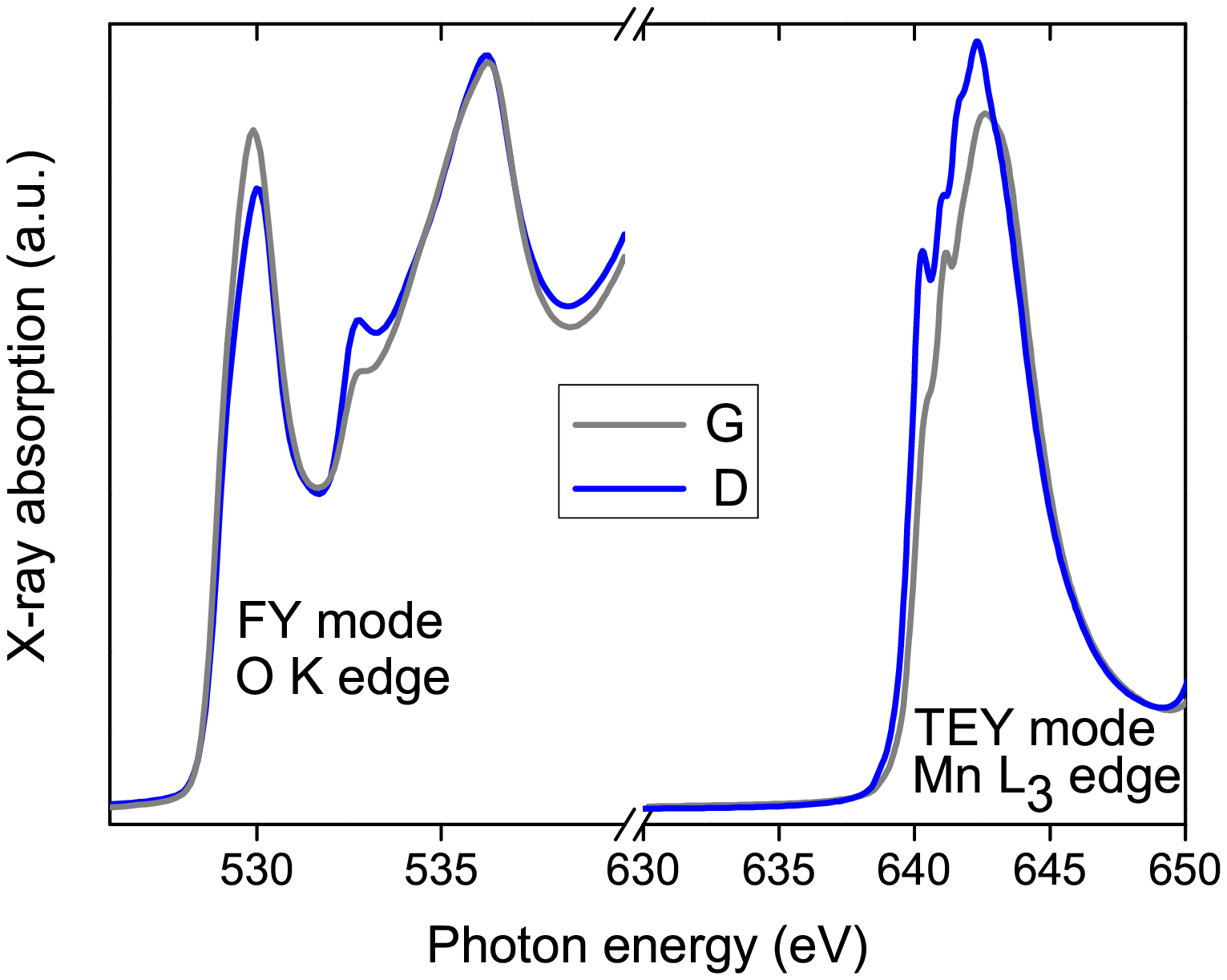}
\fi
\ifnote{\sf\textcolor{red}{[XAS-DvsG]\;}}\fi
\caption{\ifnote{\sf\textcolor{red}{[XAS-DvsG]\;}}\fi
(Color online)
XA spectra of samples D and G.
On the left side the oxygen K edge (FY mode) is shown
with the prepeak increasing with higher oxygen content.
The right side shows the manganese L$_3$ edge (TEY mode)
with different amounts of Mn$^{2+}$ for differently oxidized samples.}
%
\end{figure}

\begin{figure}
\iffig
\centering
\includegraphics[width=0.45\textwidth]{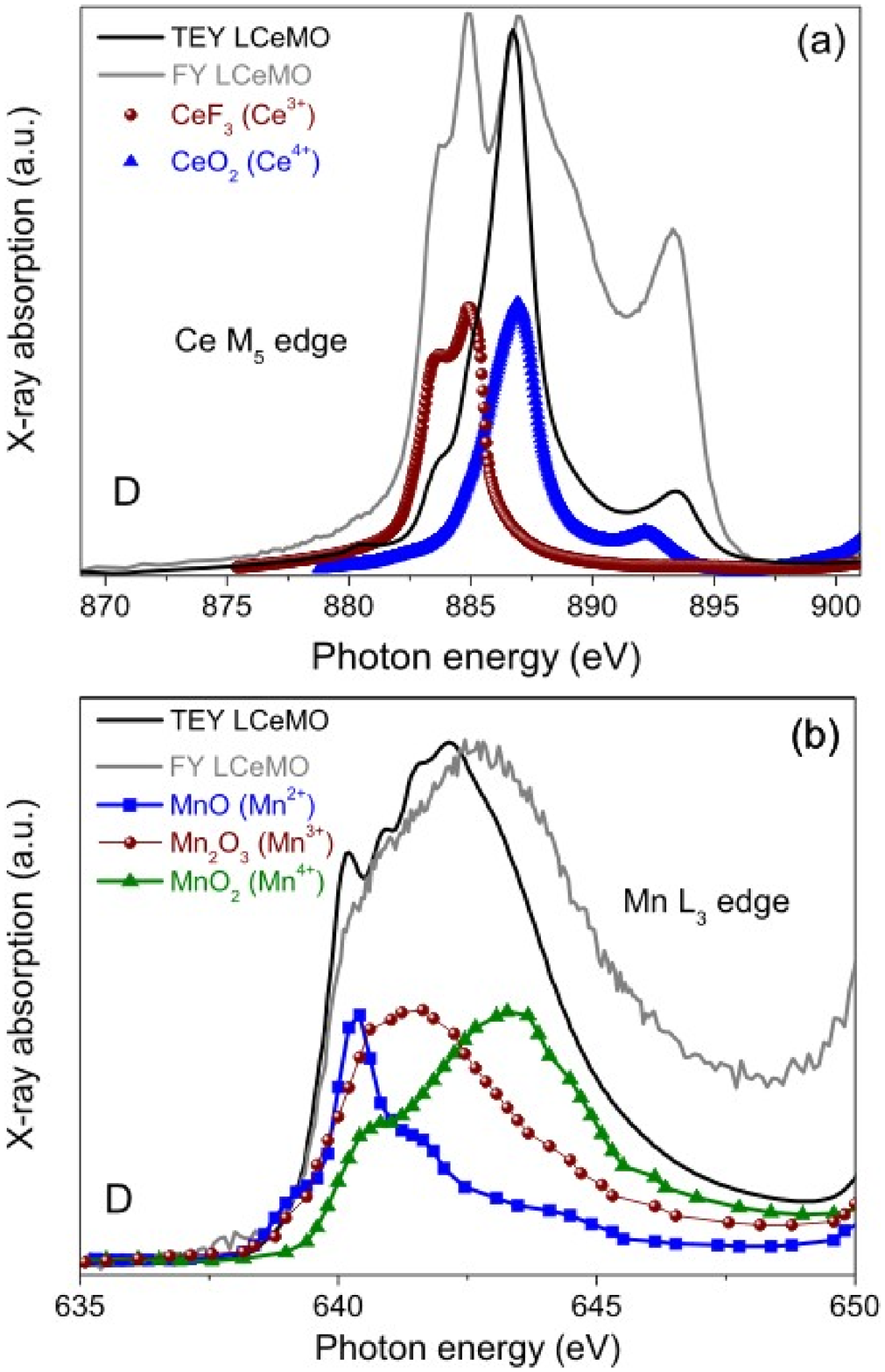}
\fi
\ifnote{\sf\textcolor{red}{[XAS-Ce-Mn]\;}}\fi
\caption{\ifnote{\sf\textcolor{red}{[XAS-Ce-Mn]\;}}\fi
(Color online)
XA spectra of sample D in TEY and FY mode (scaled to the TEY intensity)
at different absorption edges:
(a)
Cerium M$_5$ edge; reference spectra of CeO$_2$ and CeF$_3$
were added for comparison.
Please note the mixture of Ce$^{3+}$ and Ce$^{4+}$ in FY mode.
(b)
Manganese L$_3$ edge.
The FY data are self-absorption corrected following a procedure
by Ref.~[\onlinecite{Troeger92}].
Reference spectra of MnO (blue), Mn$_2$O$_3$ and MnO$_2$
were added for comparison.
Please note the missing Mn$^{2+}$ shoulder in FY mode.}
%
\end{figure}

\end{document}